\documentclass[journal]{IEEEtran}

\usepackage{cite,amssymb,amsmath}
\usepackage{color,graphicx}

\newtheorem{secthm}{Theorem}[section]
\newtheorem{seccor}[secthm]{Corollary}
\newtheorem{seclem}[secthm]{Lemma}

\newtheorem{secprop}[secthm]{Proposition}
\newtheorem{secdefn}[secthm]{Definition}
\newtheorem{secrem}[secthm]{Remark}

\newtheorem{secsasm}[secthm]{Standing Assumption}
\newtheorem{secprob}[secthm]{Problem}

\newcommand{\bA} { {\mathbb A}}

\newcommand{\bR} { {\mathbb R}}
\newcommand{\cU} { {\mathcal U}}
\def\red{\hfill $\lhd$}

\allowdisplaybreaks[4]

\title{
Controller Reduction for Nonlinear Systems by Generalized Differential Balancing 
}

\author{Yu Kawano% and Jacquelien M. A. Scherpen, {\it Senior Member, IEEE}% <-this % stops a space
\thanks{Y.~Kawano is with the Graduate School of Advanced Science and Engineering, Hiroshima University, Higashi-Hiroshima, Japan (email: ykawano@hiroshima-u.ac.jp).}
\thanks{This work was supported in part by JSPS KAKENHI Grant Numbers JP21H04875 and JP21K14185.}}

\begin{document}

\maketitle
\thispagestyle{empty}
\pagestyle{empty}

%%%%%%%%%%%%%%%%%%%%%%%%%%%%%%%%%%%%%%%%%%%%%%%%%%%%%%%%%%%%%%%%%%%%%%%%%%%%%%%%
\begin{abstract}
In this paper, we aim at developing computationally tractable methods for nonlinear model/controller reduction. Recently, model reduction by generalized differential (GD) balancing has been proposed for nonlinear systems with constant input-vector fields and linear output functions. First, we study incremental properties in the GD balancing framework. Next, based on these analyses, we provide GD LQG balancing and GD $H_\infty$-balancing as controller reduction methods for nonlinear systems by focusing on linear feedback and observer gains. Especially for GD $H_\infty$-balancing, we clarify when the closed-loop system consisting of the full-order  system and a reduced-order controller is exponentially stable. All provided methods for controller reduction can be relaxed to linear matrix inequalities.
\end{abstract}

%%%%%%%%%%%%%%%%%%%%%%%%%%%%%%%%%%%%%%%%%%%%%%%%%%%%%%%%%%%%%%%%%%%%%%%%%%%%%%%%
\section{Introduction}
An important but difficult challenge for model/controller reduction of nonlinear systems is to establish computationally tractable methods while guaranteeing stability properties of reduced-order systems/systems controlled by reduced-order controllers. Most of existing methods focus on one aspect; see e.g.~\cite{LMG:02,FT:08,HLB:12,Kashima:16,KS:21} for computational tractability and e.g.,\cite{Scherpen:93,FS:05,BVS:14,KS:17,KS:19,Astolfi:10} for theoretical analysis. A few papers~\cite{BVS:13,KS:15,SA:17,KBS:20} aim at balancing these two. It is worth emphasizing that these papers study \emph{model} reduction only. In contrast to linear systems~\cite{JS:83,OM:89,MG:91,OA:12}, \emph{controller} reduction has been rarely studied for nonlinear systems. A few papers~\cite{PF:97,YW:01} proceed with theoretical analysis based on solutions to Hamilton-Jacobi equations/inequalities (HJE/HJI). As well recognized, solving HJE/HJI is a computationally challenging problem.

As computationally tractable nonlinear \emph{model} reduction methods with theoretical analysis, generalized incremental (GI) balancing~\cite{BVS:13} and generalized differential (GD) balancing~\cite{KS:15} are proposed for nonlinear systems with constant input-vector fields and linear output functions, in the contraction~\cite{LS:98,FS:14,KBC:20,SB:04,AJP:16,Sontag:10,KCS:21,KH:21} and incremental stability~\cite{Angeli:02} frameworks, respectively. However, in GD balancing~\cite{KS:15}, the main focus is not on the original systems, but their variational systems. In other words, properties of the full/reduced-order original (i.e.  non-variational) systems are not well studied in GD balancing~\cite{KS:15}. 

Toward developing controller reduction methods for nonlinear systems in the GD balancing framework, the first objective in this paper is to study incremental stability (IS) properties~\cite{FS:14,Angeli:02} (i.e. stability properties of any pair of trajectories) for the original system based on GD balancing. We show that a system admitting a GD controllability Gramian is incrementally exponentially stable as a control system, and a similar statement holds for a GD observability Gramian. The IS properties guaranteed by this paper are preserved under model reduction by GD balanced truncation. Moreover, an upper bound on the reduction error is estimated. Note that \cite{BVS:13} for GI balancing does not study a convergence between a pair of trajectories. 

GD balancing is applicable only for stable systems. In the linear case, linear quadratic Gaussian (LQG) balanced truncation~\cite{JS:83,OM:89} is known to be a model reduction method for unstable systems. The second objective in this paper is to extend LQG balancing in the GD balancing framework; the proposed approach is called GD LQG balancing. As for linear systems, GD LQG balancing has strong connections with GD coprime factorizations proposed in this paper, as extensions of coprime factorizations. As a byproduct of GD LQG balancing analysis with GD coprime factorizations, we present an observer-based dynamic stabilizing controller while the separation principle does not hold for nonlinear systems in general. This dynamic controller raises the question of whether its reduced-order controller stabilizes the full-order system. However, the answer is not clear even in the linear case while there are researches in this direction~\cite{JS:83,OM:89}.

For stabilizing reduced-order control design, $H_\infty$-balancing has been proposed for linear systems~\cite{MG:91} and extended to nonlinear systems~\cite{PF:97,YW:01} with HJE/HJI. The last and main objective of this paper is to develop GD $H_\infty$-balancing based on GD balancing and GD LQG balancing. The main difference from~\cite{PF:97,YW:01} is that the proposed method can be achieved by solving sets of linear matrix inequalities instead of nonlinear partial differential equations/inequalities. In other words, this paper can be viewed as the first paper that provides a computationally tractable method for controller reduction of nonlinear systems while guaranteeing the stability of the closed-loop systems consisting of the full-order  systems and reduced-order controllers. Such analyzable and computationally tractable methods are developed by focusing on linear feedback and observer gains.

The remainder of this paper is organized as follows. In Section~\ref{GD:sec}, we proceed with IS analysis of GD balancing. In Sections~\ref{GDLQG:sec} and~\ref{GDH:sec}, GD LQG balancing and GD $H_\infty$-balancing are developed, respectively. Proposed methods are illustrated by means of examples in Section~\ref{Ex:sec}. Concluding remarks are given in Section~\ref{Con:sec}. All proofs are shown in Appendices.

{\it Notation}: 
The sets of real numbers and nonnegative real numbers are denoted by~$\bR$ and~$\bR_+$, respectively. For symmetric~$P \in \bR^{n \times n}$, $\lambda_{\max} (P)$ denotes its maximum eigenvalue. The inequality~$P \succ 0$~($P \succeq 0$) means that~$P$ is symmetric and positive (semi) definite. For a vector~$x \in \bR^n$, the weighted norm by a matrix $P \succ 0$ is denoted by~$|x|_P:= \sqrt{x^\top P x}$. If~$P$ is identity, this is simply denoted by~$|x|$. Also, the $\infty$-norm is denoted by~$|x|_\infty:= \max_{i =1,\dots,n} |x_i|$. For a signal~$u: \bR_+ \to \bR^m$, the $L_2$-norm is denoted by~$\|u\|:=( \int_0^{\infty} | u(t) |^2 dt)^{1/2}$. The set of signals~$u$ for which~$( \int_a^b | u(t) |^2 dt)^{1/2}$ (resp. $\sup_{t \in [a, b]} |u(t)|_\infty$) is bounded, is denoted by~$L_2^m[a,b]$ (resp. $L_\infty^m[a,b]$). We define another set of signals by $\cU:=\{ \int_0^t v(\tau) d\tau : v \in L_2^m[0, \infty) \} \cap L_\infty^m [0, \infty)$. 

\section{Generalized Differential Balancing}\label{GD:sec}
In this section, we proceed with analysis of generalized differential (GD) balancing~\cite{KS:15,KS:17}. In particular, we study incremental stability properties based on a so-called GD controllability/observability Gramian. Then, we establish the bridge between those GD Gramians and incremental energy functions. After that, we study GD balanced truncation and provide an upper error bound for model reduction.

\subsection{Generalized Differential Gramians}
Consider the system:
\begin{align}
\left\{\begin{array}{l}
\dot x = f(x) + Bu,\\
y= C x,
\end{array}\right.
\label{sys}
\end{align}
where~$x(t) \in \bR^n$,~$u(t) \in \bR^m$, and~$y(t) \in \bR^p$ denote the state, input, and output, respectively. The function~$f: \bR^n \to \bR^n$ is of class~$C^1$,~$B \in \bR^{n \times m}$, and~$C \in \bR^{p \times n}$.  Let~$\phi(t,x_0,u)$ denote the solution~$x(t)$ to the system~\eqref{sys} at time~$t \in \bR_+$ starting from~$x(0)=x_0 \in \bR^n$ with~$u(t) \in \bR^m$. Note that if~$u(\cdot)$ is continuous, the solution~$\phi(t,x_0,u)$ is a class~$C^1$ function of~$t$ and~$x_0$ as long as it exists.

The differential balancing framework is developed by using the system~\eqref{sys} and its \emph{variational system}:
\begin{align}
\left\{\begin{array}{l}
\dot {\delta x} = \partial f(x)  \delta x + B \delta u,\\
\delta y= C \delta x,
\end{array}\right.
\label{vsys}
\end{align}
where~$\delta x(t) \in \bR^n$,~$\delta u(t) \in \bR^m$, and~$\delta y(t) \in \bR^p$ denote the state, input, and output of the variational system, respectively, and $\partial f(x) := \partial f(x)/\partial x$, $x \in \bR^n$.

Using the variational system, a \emph{GD controllability Gramian} and \emph{GD observability Gramian} are respectively defined as solutions~$X \succ 0$ and~$Y \succ 0$ ($X, Y \in \bR^{n\times n}$) to the following GD Lyapunov inequalities for~$\varepsilon \ge 0$:
\begin{align}
\partial f(x) X + X \partial^\top f(x) + B B^\top + \varepsilon X \preceq 0,& \; \; \forall x \in \bR^n,
\label{Lyap_con}\\
Y \partial f(x) + \partial^\top f(x) Y + C^\top C + \varepsilon Y \preceq 0,& \; \; \forall x \in \bR^n.
\label{Lyap_ob}
\end{align}
The above inequalities look like slight extensions of the original definitions of GD Gramians~\cite{KS:15,KS:17}, where $\varepsilon = 0$ is assumed. However, our main interest is in the case $\varepsilon > 0$. In this case, we can guarantee stability properties of the system~\eqref{sys} based on the GD Gramians (i.e., based on the variational system~\eqref{vsys}), as is investigated in the next subsection. 

At present, the existence of each GD Gramian is not clear. For another GD Lyapunov inequality, \cite[Proposition 3]{AJP:16} shows that an IES system admits a matrix-valued solution. It is not yet clear when this becomes a constant. On the other hand, there can be multiple GD controllability/observability Gramians. In bounded real or positive real balancing, the corresponding Riccati \emph{equations} admit multiple solutions, and the concept of the minimal and maximal solutions have been introduced \cite{GA:04}. However, for the \emph{inequalities}, the minimal or maximal solutions may not exist because the inequality relaxation makes the class of solutions wider, which allows us to impose additional requirements for model reduction such as structure-preservation; see e.g. Remark~\ref{rem:structure} below.

\subsection{Incremental Stability Analysis}
In the original work for differential balancing, properties of variational systems are investigated~\cite{KS:15,KS:17}. As known in contraction analysis~\cite{FS:14,KBC:20}, properties of variational systems have strong connections with \emph{incremental} properties, i.e., system properties stated in terms of a pair of trajectories. In this section, we investigate incremental stability (IS) by using GD Gramians. To this end, we use the auxiliary system~\cite{Angeli:02} consisting of the system~\eqref{sys} and its copy:
\begin{align}
\left\{\begin{array}{ll}
\dot x = f(x) + Bu, \\
\dot x' = f(x') + B u',\\
y= C x, \\
y'= C x'.
\end{array}\right.
\label{auxsys}
\end{align}

First, we recall the definition of incremental exponential stability as a property of trajectories~$(x(t),x'(t))$ of the auxiliary system for the same input~$u(t) = u'(t)$.
\begin{secdefn}[\!\!\cite{FS:14,Angeli:02}]
The system~\eqref{sys} is said to be \emph{incrementally exponentially stable} (IES) (with respect to~$u(t) \in \bR^m$) if~$\phi(t,x_0,u) \in \bR^n$ for each $x_0\in\bR^n$ and $t\in\bR_+$,
and there exist~$k, \lambda > 0$ such that 
\begin{align*}
| \phi (t,x_0,u) - \phi (t,x'_0,u) | \le k e^{-\lambda t} |x_0 - x'_0|, \; \forall t\in\bR_+,
\end{align*}
for any~$(x_0,x'_0) \in \bR^n \times \bR^n$.
\red
\end{secdefn}

Now, we are ready to state the first result. Thanks to newly introduced~$\varepsilon>0$, IS properties can be studied by using the GD controllability Gramian, differently from the existing works for GD balancing~\cite{KS:15,KS:17}.
\begin{secthm}\label{thm:Lyap_con}
Given~$\varepsilon > 0$, if the system~\eqref{sys} has a GD controllability Gramian~$X \succ 0$, then
\begin{enumerate}
\item $|\phi(t,x_0, u)-\phi(t,x'_0, u')|$, $t \in \bR_+$ exists for any~$(x_0,x'_0) \in \bR^n \times \bR^n$ and~$(u,u')$ such that $u-u' \in L_2^m [0,\infty)$ and~$u' \in \cU$;
\item $\lim_{t \to \infty} |\phi(t,x_0,u) -\phi(t,x'_0,u')|=0$ for any $(x_0, x'_0) \in \bR^n \times \bR^n$ and~$(u,u') \in \cU \times \cU$ such that $u-u' \in L_2^m[0,\infty)$; 
\item the system is IES with respect to each~$u \in \cU$;
\item for each constant input~$u = u^* \in \bR^m$, the system has the globally exponentially stable (GES) equilibrium point~$x^* \in \bR^n$;
\item for each~$x_0 \in \bR^n$ and~$u \in L_2^m[0, \infty)$, $|\phi(t, x_0, u)|$ is a bounded function of~$t \in \bR_+$.
\red
\end{enumerate}
\end{secthm}
\begin{secrem}
As can be noticed from the proof of Theorem~\ref{thm:Lyap_con}, the condition $u \in \cU$ (resp. $u' \in \cU$) can be replaced with $u \in L_2^m[0, \infty)$ (resp. $u' \in L_2^m[0, \infty)$) in items 1) -- 3).
\red
\end{secrem}

We have a similar result for observability.
\begin{secthm}\label{thm:Lyap_ob}
Given~$\varepsilon > 0$, if the system~\eqref{sys} has a GD observability Gramian~$Y \succ 0$, then
\begin{enumerate}
\item $|\phi(t,x_0, u^*)-\phi(t,x'_0, u^*)|$,~$t\in\bR_+$ exists for any~$(x_0,x'_0) \in \bR^n \times \bR^n$ and $u^* \in \bR^m$;
\item for each constant input~$u = u^* \in \bR^m$, the system is IES and has the GES equilibrium point~$x^* \in \bR^n$.
 \red
\end{enumerate}
\end{secthm}

\begin{secrem}
In contraction analysis of autonomous systems, a considered set ($\bR^n$ in this case) is typically assumed to be positively invariant, see e.g.,~\cite{FS:14,SB:04,AJP:16}. However, we do not assume that~$\bR^n$ is robustly positively invariant~\cite{BM:08} (positive invariance as a control system) in Theorems~\ref{thm:Lyap_con} and~\ref{thm:Lyap_ob}.  Namely, IES is guaranteed without any assumption for positive invariance. Furthermore, we can conclude that the system has a unique equilibrium point for each constant input, which is also a new finding of this paper. These are benefits of newly introduced~$\varepsilon > 0$. In fact, this is the first paper studying IES in the balancing framework. In particular, there is no study for IS properties based on a controllability Gramian other than Theorem~\ref{thm:Lyap_con}; even the paper~\cite{BVS:13} on generalized incremental balancing has not studied IS properties. Only a statement similar to that of item~1) in Theorem~\ref{thm:Lyap_ob} for an observability Gramian is found in~\cite[Lemma~2]{BVS:13}. In other words, item 2) is also a new finding of this paper. 
\red
\end{secrem}

\begin{secrem}\label{rem:epsilon}
Even when~$\varepsilon = 0$, item 1) of each Theorems~\ref{thm:Lyap_con} and~\ref{thm:Lyap_ob} can be shown. Moreover, the existence of a GD observability Gramian~$Y \succ 0$ for~$\varepsilon = 0$ implies 
\begin{align*}
\lim_{t \to \infty}|y(t)- y'(t)|=0
\end{align*}
for each constant input~$u= u' = u^*$. In addition, if the system has the detectability property:
\begin{align}
&\lim_{t \to \infty} | u(t) - u'(t) |  = 0, \; \lim_{t \to \infty} |y(t) -y'(t) | =0 \nonumber\\
&\hspace{5mm}\implies \lim_{t \to \infty} |\phi(t,x_0, u)-\phi(t,x'_0, u')| =0,
\label{detectability}
\end{align}
then $\lim_{t \to \infty} |\phi(t,x_0, u)-\phi(t,x'_0, u')| =0$ is guaranteed. 
\red
\end{secrem}

The inequality~\eqref{Lyap_con} or~\eqref{Lyap_ob} is not only helpful for studying stability properties, but also characterizes structures of the mapping~$f$ itself.
\begin{seccor}\label{cor:mapping}
Given~$\varepsilon > 0$, if either~\eqref{Lyap_con} or~\eqref{Lyap_ob} holds, then $f:\bR^n \to \bR^n$ is a globally diffeomorphism. 
\red
\end{seccor}

\subsection{Connections with Incremental Energy Functions}
Linear/nonlinear balancing has a strong connection with the so-called controllability and observability functions~\cite{Antoulas:05,Scherpen:93}.  By using the auxiliary system, define the following incremental controllability and observability functions, respectively,
\begin{align}
J_u (x_0,x'_0) :=  \inf_{\substack{u-u' \in L_2^m[0,\infty) \\ (x(0),x'(0)) = (x_0,x'_0) \\ x(-\infty)=x'(-\infty)}} \int_{-\infty}^0 |u(t) - u'(t)|^2 dt, \label{con_func}
\end{align}
and
\begin{align}
J_y (x_0,x'_0,u^*) &:= \int_0^\infty |y(t) - y'(t)|^2 dt, \label{ob_func}\\
& (x(0),x'(0))=(x_0,x'_0 ), \; u=u' = u^* \in \bR^m. \nonumber
\end{align}
\begin{secrem}
The incremental controllability and observability functions (\ref{con_func}) and (\ref{ob_func}) are slightly different from those in \cite{BVS:14}. When compared to the incremental controllability function based on the sum~$u + u'$ in \cite[Definition 6]{BVS:14}, (\ref{con_func}) can be regarded in a more natural \emph{incremental} function as the difference~$u-u'$ is considered (and we do not require additionally $x(-\infty) = x'(-\infty) = 0$).
The incremental observability function in \cite[Definition 5]{BVS:14} is defined as the supremum over all input functions.
\red
\end{secrem}

A GD controllability/observability Gramian naturally provides a lower/upper bound on these incremental energy functions, as is stated next.
\begin{secthm}\label{thm:con_lowbd}
Consider the system~\eqref{sys} and its auxiliary system~\eqref{auxsys}. Suppose that
\begin{enumerate}
\renewcommand{\theenumi}{\Roman{enumi}}
\item $J_u (x_0,x'_0)$ in~\eqref{con_func} exists for any~$(x_0,x'_0) \in \bR^n \times \bR^n$;
\item given~$\varepsilon \ge 0$, a GD controllability Gramian~$X \succ 0$ exists.
\end{enumerate}
Then,
\begin{align}
|x_0 - x'_0|_{X^{-1}}^2 \le J_u (x_0,x'_0) \label{con_func_lb}
\end{align}
for any~$(x_0,x'_0) \in \bR^n \times \bR^n$.
\red
\end{secthm}

\begin{secthm}\label{thm:ob_upbd}
Consider the system~\eqref{sys} and its auxiliary system~\eqref{auxsys}. Suppose that
\begin{enumerate}
\renewcommand{\theenumi}{\Roman{enumi}}
\item $J_y (x_0,x'_0,u^*)$ in~\eqref{ob_func} exists for some constant input~$u^* \in \bR^m$ and for any~$(x_0,x'_0) \in \bR^n \times \bR^n$;
\item given~$\varepsilon \ge 0$, a GD observability Gramian~$Y \succ 0$ exists;
\item the system~\eqref{sys} has the detectability property~\eqref{detectability}.
\end{enumerate}
Then,
\begin{align}
J_y (x_0,x'_0,u^*) \le |x_0 - x'_0|_Y^2 \label{ob_func_ub}
\end{align}
for any~$(x_0,x'_0) \in \bR^n \times \bR^n$.
\red
\end{secthm}

\subsection{Model Reduction}\label{sec:mod}
GD balanced truncation has been developed based on the following proposition. Originally, this is shown for~$\varepsilon = 0$ \cite[Theorem~19.17]{KS:15}, but holds for each~$\varepsilon \ge 0$.
\begin{secprop}
Given $\varepsilon \ge 0$, if a system~\eqref{sys} has both GD controllability and observability Gramians~$X \succ 0$ and~$Y\succ 0$, then there exist \emph{GD balanced coordinates} in which
\begin{align*}
X = Y = \Sigma := {\rm diag} \{ \sigma_1, \dots, \sigma_n \}, \; \sigma_1 \ge  \cdots \ge \sigma_n > 0 
\end{align*}
hold. \red
\end{secprop}

A change of coordinates~$z = Tx$ into GD balanced coordinates can be constructed similarly as for balanced coordinates between the controllability and observability Gramians of linear systems~\cite{Moore:81,Antoulas:05}. Hereafter in this subsection, suppose that the system~\eqref{sys} has both positive definite GD controllability and observability Gramians~$X$ and~$Y$. For the sake of simplicity of the description, we further suppose that a realization~\eqref{sys} is GD balanced, namely~$X=Y=\Sigma$. 

For~$r$ satisfying~$\sigma_r>\sigma_{r+1}$, define
\begin{align}
\Sigma_1:=  {\rm diag} \{ \sigma_1,\dots,\sigma_r\}, \;
\Sigma_2:=  {\rm diag} \{ \sigma_{r+1},\dots,\sigma_n\}.
\end{align}
Correspondingly, we divide the system~\eqref{sys} into
\begin{align}
&x = \begin{bmatrix} x_1 \\ x_2 \end{bmatrix}, \;
f(x) = \begin{bmatrix}  f_1(x_1,x_2) \\ f_2(x_1,x_2) \end{bmatrix}, \;
B = \begin{bmatrix} B_1 \\ B_2 \end{bmatrix}, \nonumber\\
&C =\begin{bmatrix} C_1 & C_2 \end{bmatrix}.
\label{sys_div}
\end{align}
A reduced-order model by GD balancing is constructed as
\begin{align}
\left\{\begin{array}{l}
\dot x_r =  f_1(x_r + c_1,c_2) + B_1 u,\\
y_r= C_1 (x_r + c_1) + C_2 c_2,
\end{array}\right.
\label{rsys}
\end{align}
where~$c :=[c_1^\top c_2^\top]^\top  \in \bR^n$ is arbitrary. 

In the balanced coordinates, $\Sigma$ is both GD controllability and observability Gramians. Namely, the system~\eqref{sys_div} satisfies~\eqref{Lyap_con} and~\eqref{Lyap_ob} for~$X=Y=\Sigma$. Now, the GD Lyapunov inequality~\eqref{Lyap_con} becomes
\begin{align*}
&\begin{bmatrix} 
\partial_{x_1} f_1  & \partial_{x_2} f_1 \\ 
\partial_{x_1} f_2 & \partial_{x_2} f_2
\end{bmatrix}
\begin{bmatrix} \Sigma_1 & 0 \\ 0 & \Sigma_2 \end{bmatrix}
+
\begin{bmatrix} \Sigma_1 & 0 \\ 0 & \Sigma_2 \end{bmatrix}
\begin{bmatrix} 
\partial_{x_1} f_1  & \partial_{x_2} f_1 \\ 
\partial_{x_1} f_2 & \partial_{x_2} f_2
\end{bmatrix}^\top\\
&+ \begin{bmatrix} B_1 B_1^\top & B_1 B_2^\top \\ B_1 B_2^\top & B_2 B_2^\top\end{bmatrix} + \varepsilon \begin{bmatrix} \Sigma_1 & 0 \\ 0 & \Sigma_2 \end{bmatrix} \preceq 0,
\end{align*}
where $\partial_{x_i} f_j := \partial f_j/\partial x_i$, $i,j=1,2$. From the $(1,1)$-th block diagonal element, one notices that~$\Sigma_1 \succ 0$ is a GD controllability Gramian for the reduced-order model~\eqref{rsys}. A  similar discussion holds for a GD observability Gramian. Therefore, similar statements as Theorems~\ref{thm:Lyap_con} and~\ref{thm:Lyap_ob} hold for the reduced-order model. That is, GD balanced truncation preserves stability properties.

\begin{secrem}\label{rem:structure}
As mentioned in the previous subsection, there can be multiple GD Gramians. This is beneficial, for instance, for structure-preserving model reduction. Consider the interconnection of two subsystems:
\begin{align*}
\left\{\begin{array}{l}
\dot x_1 = f_1(x_1) + g_1(x_2) + B_1 u_1,\\
\dot x_2 = g_2(x_1) + f_2(x_2) + B_2 u_2,\\
y_1 = C_1 x_1, \; y_2 = C_2 x_2.
\end{array}\right.
\end{align*}
If we specify the structures of GD Gramians $X$ and $Y$ as block-diagonal matrices, i.e. $X= {\rm diag}\{ X_1, X_2\}$ and $Y= {\rm diag}\{ Y_1, Y_2\}$ with the compatible dimensions, then we can achieve model reduction of each subsystem separately, which preserves the interconnection structure. 
\red
\end{secrem}

An error between the full-order  system and reduced-order model can be estimated for a specific drift vector field. In fact, a similar error bound can be established by combining~\cite[Theorem 3]{BVS:14} and~\cite[Theorem 18]{BVS:14}. However, our result is more general, and the proof is used later in analysis of controller reduction. 
\begin{secthm}\label{thm:error}
Suppose that
\begin{enumerate}
\renewcommand{\theenumi}{\Roman{enumi}}
\item given~$\varepsilon \ge 0$, the system~\eqref{sys} has both GD controllability and observability Gramians~$X \succ 0$ and~$Y \succ 0$; 
\item $\sigma_r>\sigma_{r+1}$;
\item in GD balanced coordinates, there exists~$c \in \bR^n$ such that~$f(x_r+c_1,c_2)= -f(-x_r + c_1,c_2)$ for all $x_r \in \bR^r$.
\end{enumerate}
Then, the outputs of the system~\eqref{sys} and its reduced-order model~\eqref{rsys} satisfy
\begin{align}
\| y - y_r\| \le 2 \sum_{i=r+1}^n \sigma_i \|u\|.
\label{error}
\end{align}
for any~$u \in L_2^m[0, \infty)$ under the initial states~$x(0)=c$ and~$x_r(0)=0$. \red
\end{secthm}
\begin{secrem}\label{rem:c}
In Theorem~\ref{thm:error}, if~$\varepsilon > 0$, the constant~$c$ can be chosen as zero without loss of generality. First, item~4) of Theorem~\ref{thm:Lyap_con} for~$u=0$ implies that there exists a unique equilibrium~$x^*$. Next, for~$x_r=0$, the condition in item~III) of Theorem~\ref{thm:error} becomes~$2f(c_1,c_2) = 0$, which implies~$c = x^*$.  In the new coordinates~$\hat x := x - x^*$, the equilibrium point is zero, and the condition in item~III) of Theorem~\ref{thm:error} holds for~$c = 0$. On the other hand, in the~$\hat x$-coordinates, the output becomes an affine function~$y = C (\hat x + x^*)$. However, one can define the new outputs~$\hat y := y - C x^* = C\hat x$. Note that the GD Gramians do not depend on the shifts of coordinates~$\hat x = x - x^*$ and outputs~$\hat y = y - C x^*$, since the variational system~\eqref{vsys} is independent from them. 
\red
\end{secrem}
	
Hereafter, to reduce the complexity of discussions, we impose the following without loss of generality.
\begin{secsasm}\label{asm:eq}
$f(0)=0$ in~\eqref{sys} and~$c = 0$ in~\eqref{rsys}.
\red
\end{secsasm}

\begin{secrem}\label{rem:local}
Under $f(0)=0$, if~\eqref{Lyap_con} has a solution~$X \succ 0$ on a convex set~$D\subset \bR^n$ containing the origin, then any trajectory starting from~$D$ with the zero input converges to the origin. More generally, as long as~$\phi(t,x_0,u) \in D$, $t \in \bR_+$, the trajectory converges to the origin for any~$x_0 \in D$ and~$u \in \cU$. Therefore, the obtained results for GD balancing in this section can be extended to model reduction on $D$.
\red
\end{secrem}

Under Assumption~\ref{asm:eq}, item~III) of Theorem~\ref{thm:error} requires that~$f$ is an odd function. That is, each $f_i$, $i=1,\dots,n$ is odd. Odd nonlinearities naturally appear in physical systems such as nonlinear pendulums~\cite{Khalil:02}, ball and beam systems~\cite{HSK:92}, mass-spring-damper systems with hardening springs~\cite{Khalil:02}, and van der Pol oscillators (or more generally negative resistance oscillators)~\cite{Khalil:02}. Moreover, relays and saturations, standard static nonlinearities~\cite{Khalil:02} can be approximated by $\arctan (x)$ that is an odd function.
 
Note that we assume $f$ to be odd in error analysis of model reduction and later in controller reduction to guarantee the stability of the closed-loop system with a reduced-order controller. Otherwise, this assumption is not required. In fact, \eqref{Lyap_con} and \eqref{Lyap_ob} can have solutions without this assumption. Consider the following scaler system: $\dot x = -x -x^2-x^3 +u$, where $f$ is not odd. One can confirm that $X=2$ is a solution to \eqref{Lyap_con} for~$\varepsilon = 0.1$.

%%%%%%%%%%%%%%%%%%%%%%%%%%%%%%%%%%%%%%%
%%%%%%%%%%%%%%%%%%%%%%%%%%%%%%%%%%%%%%%
\section{Generalized Differential LQG Balancing} \label{GDLQG:sec}
As balanced truncation for linear systems is applicable only for stable systems, GD balancing requires that the system~\eqref{sys} has IS properties. For unstable linear/nonlinear systems, so-called LQG balancing has been proposed~\cite{JS:83,SS:94,OM:89}. In this section, to deal with non-IES systems, we develop a new LQG balancing method in the GD balancing framework, which we call \emph{GD LQG balancing}. Next, inspired by results on LQG balancing, we also propose \emph{GD coprime factorizations} and investigate its GD balancing in terms of GD LQG balancing of the system~\eqref{sys}. Based on the results on coprime factorizations, we study stabilizing dynamic controller design. 

\subsection{Generalized Differential Riccati Equations}
Motivated by LQG balancing~\cite{JS:83,SS:94,OM:89}, we consider the following past energy function,
\begin{align}
J_- (x_0,x'_0)
= \inf_{\substack{u-u' \in L_2^m(-\infty,0] \\ (x(0),x'(0)) = (x_0,x'_0) \\ x(-\infty) = x'(-\infty)}} \int_{-\infty}^0 \eta (t) dt, \label{cost-}
\end{align}
and post energy function,
\begin{align}
J_+ (x_0,x'_0)
 = \inf_{\substack{u-u' \in L_2^m[0,\infty) \\ (x(0),x'(0)) = (x_0,x'_0) \\ x(\infty)=x'(\infty)}} \int_0^{\infty} \eta (t) dt, \label{cost+}
\end{align}
where $\eta (t) := |y (t) - y' (t) |^2 + | u(t) - u' (t) |^2$.

As done for the incremental controllability/observability function in the previous section, we first estimate their lower/upper bound by using each solution $P\succ 0$ or $Q \succ 0$  ($P$, $Q \in \bR^{n\times n}$) to the following \emph{GD control Riccati inequality} or \emph{GD filter Riccati inequality} for $\varepsilon \ge 0$:
\begin{align}
P \partial f(x) + \partial^\top f(x) P &- P B B^\top P \nonumber\\
&+ C^\top C + \varepsilon P \preceq 0, \; \forall x \in \bR^n,
\label{Ric_con}\\
\partial f(x) Q + Q \partial^\top f(x) &- Q C^\top C Q\nonumber\\
&+ B B^\top + \varepsilon Q \preceq 0, \; \forall x \in \bR^n. \label{Ric_ob}
\end{align}

An upper bound on~$J_+ (x_0,x'_0)$ can be estimated by using the GD control Riccati inequality~\eqref{Ric_con} as stated below.
\begin{secthm}\label{thm:up_post}
For the system~\eqref{sys} and its auxiliary system~\eqref{auxsys}, suppose that
\begin{enumerate}
\renewcommand{\theenumi}{\Roman{enumi}}
\item $J_+ (x_0,x'_0)$ in~\eqref{cost+} exists for any~$(x_0,x'_0) \in \bR^n \times \bR^n$; 
\item given~$\varepsilon \ge 0$, the GD control Riccati inequality~\eqref{Ric_con} admits a solution~$P \succ 0$;
\item the system has the detectability property~\eqref{detectability}.
\end{enumerate}
Then,
\begin{align}
J_+ (x_0,x'_0) \le |x_0 - x'_0|_P^2 \label{cost+_ub}
\end{align}
for any~$(x_0,x'_0) \in \bR^n \times \bR^n$.
\red
\end{secthm}

A similar result is obtained for a lower bound on~$J_- (x_0,x'_0)$ and the GD filter Riccati inequality~\eqref{Ric_ob} as stated below.
\begin{secthm}\label{thm:low_past}
For the system~\eqref{sys} and its auxiliary system~\eqref{auxsys}, suppose that
\begin{enumerate}
\renewcommand{\theenumi}{\Roman{enumi}}
\item $J_- (x_0,x'_0)$ in~\eqref{cost-} exists for any~$(x_0,x'_0) \in \bR^n \times \bR^n$; 
\item given~$\varepsilon \ge 0$, the GD filter Riccati inequality~\eqref{Ric_ob} admits a solution~$Q \succ 0$. \end{enumerate}
Then 
\begin{align}
|x_0 - x'_0|_{Q^{-1}}^2 \le J_- (x_0,x'_0)
\label{cost-_lb}
\end{align}
for any~$(x_0,x'_0) \in \bR^n \times \bR^n$.
\red
\end{secthm}

If GD Riccati inequalities~\eqref{Ric_con} and~\eqref{Ric_ob} have solutions~$P \succ 0$ and~$Q \succ 0$, respectively, then as for GD Gramians~$X,Y\succ 0$, there exist coordinates, named \emph{GD LQG balanced coordinates}, in which
\begin{align}
P = Q = \Pi := {\rm diag}\{\pi_1,\dots,\pi_n\}, \label{LQGbal}\\
\pi_1 \ge \cdots \ge \pi_n > 0. \nonumber
\end{align}
Let~$\pi_r > \pi_{r+1}$, and define~$\Pi_1 := {\rm diag}\{\pi_1,\dots,\pi_r\}$. Then, in the GD LQG balanced coordinates, model reduction of the system~\eqref{sys} can be achieved in a similar manner as GD balancing. Furthermore,~$\Pi_1$ satisfies both GD Riccati inequalities~\eqref{Ric_con} and~\eqref{Ric_ob} for the reduced-order model~\eqref{rsys}. 

Finally, it is remarked that the GD controllability and observability Gramians~$X$ and~$Y$ satisfy the GD Riccati inequalities~\eqref{Ric_ob} and~\eqref{Ric_con}, respectively. This implies that GD LQG balancing can be applied to a wider class of systems than GD balancing.

\subsection{Generalized Differential Right Coprime Representations}\label{sec:rcr}
LQG balancing has been  understood as balanced truncation of the so-called right coprime representation, i.e. the closed-loop system designed by the stabilizing solution to the control Riccati equation~\cite{SS:94,OM:89}. Motivated by this, we define a \emph{GD right coprime representation} (RCR) with a solution~$P \succ 0$ to the GD control Riccati inequality~\eqref{Ric_con} as follows:
\begin{align}
\left\{\begin{array}{l}
\dot x = f(x) - B B^\top P x + B v,\\
\begin{bmatrix}
y\\u 
\end{bmatrix}
= 
\begin{bmatrix}
C \\ - B^\top P 
\end{bmatrix} x
+
\begin{bmatrix}
0 \\  I_m
\end{bmatrix}  v,
\end{array}\right.
\label{rcrsys}
\end{align}
where~$v (t)\in \bR^m$ is the new input. 

Similarly to the linear case, both GD controllability and observability Gramians for the GD RCR are constructed from solutions to the GD Riccati inequalities~\eqref{Ric_con} and~\eqref{Ric_ob} as stated below.
\begin{secthm}\label{thm:rcrsys_Gram}
Consider the system~\eqref{sys}. Suppose that given~$\varepsilon \ge 0$, the GD Riccati inequalities~\eqref{Ric_con} and~\eqref{Ric_ob} admit solutions~$P \succ 0$ and~$Q \succ 0$, respectively. Then, $(P + Q^{-1})^{-1}$ (resp. $P$) is a GD controllability (resp. observability) Gramian of the GD RCR~\eqref{rcrsys}.
\red
\end{secthm}

\begin{secrem}\label{rem:rcr}
Theorem~\ref{thm:rcrsys_Gram} implies that the GD RCR~\eqref{rcrsys} satisfies similar statements as Theorems~\ref{thm:Lyap_con} and~\ref{thm:Lyap_ob} for stability. That is, $u = - B^\top P x$ is a stabilizing controller of the system~\eqref{sys} if~$\varepsilon > 0$. In fact, in Theorem~\ref{thm:up_post}, the detectability property is not necessarily if~$\varepsilon > 0$.
\red
\end{secrem}

From Theorem~\ref{thm:rcrsys_Gram}, GD LQG balanced coordinates of the system~\eqref{sys} are also GD balanced coordinates of the GD RCR~\eqref{rcrsys}. Indeed, from~\eqref{LQGbal}, it follows that
\begin{align*}
&(P + Q^{-1})^{-1} = (\Pi + \Pi^{-1})^{-1}\\
& = {\rm diag}\{\pi_1/(1+\pi_1^2), \dots, \pi_n/(1+\pi_n^2)\}.
\end{align*} 
Therefore, in GD LQG balanced coordinates, the GD controllability Gramian~$(\Pi + \Pi^{-1})^{-1}$ and GD observability Gramian~$\Pi$ of the GD RCR~\eqref{rcrsys} are diagonal. In these coordinates, one can achieve GD balanced truncation of the GD RCR. A reduced-order model of the GD RCR is a GD RCR of the reduced-order model~\eqref{rsys}, i.e.,  a constructing reduced order model and computing a GD RCR are commutative. Furthermore, an error bound for GD balanced truncation of the GD RCR can be estimated as in Theorem~\ref{thm:error}.

\subsection{Generalized Differential Left Coprime Representations}\label{sec:GDLCR}
Next, we consider an extension of the left coprime representation (LCR)~\cite{SS:94,OM:89}, which is defined by using a solution~$Q \succ 0$ to the GD Riccati inequality~\eqref{Ric_ob} as follows:
\begin{align}
\left\{\begin{array}{l}
\dot x = f(x) -  Q C^\top C x + 
\begin{bmatrix}
B & Q C^\top
\end{bmatrix}
\begin{bmatrix}
u \\ y
\end{bmatrix},\\
z = C x +
\begin{bmatrix}
0 & - I_p
\end{bmatrix}
\begin{bmatrix}
u \\ y
\end{bmatrix},
\end{array}\right.
\label{lcrsys}
\end{align}
where~$z(t) \in \bR^p$ is the new output.

As for the GD RCR, both GD controllability and observability Gramians of the GD LCR are obtained from the GD Riccati inequalities~\eqref{Ric_con} and~\eqref{Ric_ob} as follows.
\begin{secthm}\label{thm:lcrsys_Gram}
Consider the system~\eqref{sys}. Suppose that given~$\varepsilon \ge 0$, the GD Riccati inequalities~\eqref{Ric_con} and~\eqref{Ric_ob} admit solutions~$P \succ 0$ and~$Q \succ 0$, respectively. Then, $Q$ (resp.~$(Q + P^{-1} )^{-1}$) is a GD controllability (resp. observability) Gramian of the GD LCR~\eqref{lcrsys}.
\red
\end{secthm}
 
GD LQG balanced coordinates of the system~\eqref{sys} are also GD balanced coordinates for the GD LCR~\eqref{lcrsys}. Therefore, one can achieve GD balanced truncation, and the obtained reduced-order model is a GD LCR of the reduced-order model~\eqref{rsys}. 

As briefly mentioned in Remark~\ref{rem:rcr}, the GD RCR~\eqref{rcrsys} corresponds to the closed-loop system with a stabilizing controller~$u = - B^\top P x$. In fact, the GD LCR~\eqref{lcrsys} corresponds to observer dynamics. Consider the following system having the same dynamics as the GD LCR:
\begin{align}
\dot {\hat x} = f (\hat x) -  Q C^\top C \hat x + 
\begin{bmatrix}
B & Q C^\top
\end{bmatrix}
\begin{bmatrix}
u \\ y
\end{bmatrix}.
\label{observer}
\end{align}
If its initial state is the same as the initial state of the system~\eqref{sys}, i.e.,~$\hat x(0) = x_0$, then~$\hat x = x$. Therefore, if~\eqref{observer} is IES with respect to the external inputs~$(u,y)$, then~$\hat x(t) - x(t) \to 0$ as~$t \to \infty$ for arbitrary $x_0$, $\hat x(0) \in \bR^n$, which implies that~\eqref{observer} is an observer of~\eqref{sys}. If~$\varepsilon > 0$, the IES of~\eqref{observer} can be shown by combining Theorems~\ref{thm:Lyap_con} and~\ref{thm:lcrsys_Gram}.

\subsection{Dynamic Stabilizing Controllers}
As mentioned in the previous subsections, the GD RCR and LCR correspond to a static stabilizing controller and observer, respectively. Thus, it is expected that a dynamic stabilizing controller can be constructed by combining them. In this subsection, we show that this is true even though the separation principle does not hold for nonlinear systems in general.

As a dynamic controller, we consider the following observer based stabilizing controller:
\begin{align}
\left\{\begin{array}{l}
\dot x_c = f(x_c) + B u -  Q C^\top (C x_c - y),\\
u = - B^\top P x_c.
\end{array}\right.
\label{con}
\end{align}
The closed-loop system consisting of~\eqref{sys} and~\eqref{con} is
\begin{align}
\left\{\begin{array}{l}
\dot x = f(x) - B B^\top P x_c,\\
\dot x_c = f(x_c)  - B B^\top P x_c -  Q C^\top C (x_c - x).
\end{array}\right.
\label{clsys}
\end{align}
This closed-loop system is stable, which is stated formally.
\begin{secthm}\label{thm:clstab}
Consider the system~\eqref{sys}. Suppose that given~$\varepsilon > 0$, the GD Riccati inequalities~\eqref{Ric_con} and~\eqref{Ric_ob} admit solutions~$P \succ 0$ and~$Q \succ 0$, respectively. Then, the closed-loop system~\eqref{clsys} is GES at the origin  (recall Assumption~\ref{asm:eq}).~\red
\end{secthm}

Since~$\Pi_1$ satisfies both GD Riccqti inequalities~\eqref{Ric_con} and~\eqref{Ric_ob} for the reduced-order model~\eqref{rsys} (constructed by GD LQG balancing), the following controller: 
\begin{align*}
\left\{\begin{array}{l}
\dot x_{cr} = f_1(x_{cr},0) - B_1 u -  \Pi_1 C_1^\top (C_1 x_{cr} - y_r),\\
u = - B_1^\top \Pi_1 x_{cr},
\end{array}\right.
\end{align*}
is a dynamic stabilizing controller of the reduced-order model~\eqref{rsys} with~$c=0$. Then, one may ask whether this reduced-order controller stabilizes the full-order system~\eqref{sys}. However, the answer is not very clear even in the linear case while there are researches in this direction, e.g.,~\cite{JS:83,OM:89}. Toward designing a reduced-order stabilizing controller, in the next section, we develop another balancing method as an extension of~$H_\infty$-balancing~\cite{MG:91}.

\section{Generalized Differential $H_\infty$-Balancing}\label{GDH:sec}
In this section, we extend the so-called~$H_\infty$-balancing in the GD balancing framework. The objective is to design a reduced-order controller which stabilizes the system~\eqref{sys}. We first design a full-order dynamic stabilizing controller and then consider reducing its dimension while guaranteeing closed-loop stability.

\subsection{Full-Order Dynamic Controller Design}
Now, we consider the generalized plant corresponding to the system~\eqref{sys}:
\begin{align}
G: \left\{\begin{array}{l}
\dot x = f(x) + Bu + B w_u,\\
z_u = u,\\
z_y = Cx,\\
y= C x + w_y,
\end{array}\right.
\label{gsys}
\end{align}
where~$w (t):= [w_u^\top (t) \; w_y^\top(t)]^\top \in \bR^{m+p}$ and~$z := [z_u^\top (t)\; z_y^\top (t)]^\top \in \bR^{m+p}$.

The first goal is to design an output feedback controller solving a standard~$L_2$ ($H_\infty$) control problem~\cite{DGK:89,LD:94}.
\begin{secprob}\label{prob:Hcon}
For the system~\eqref{gsys}, find a controller such that
\begin{enumerate}
\item the~$L_2$-gain of the closed-loop system from~$w$ to~$z$ is not greater than a prescribed value~$\gamma > 1$, namely~$\|z\| \le \gamma \|w \|$ for the zero initial state;
\item the closed-loop system is GES at the origin when~$w = 0$ (recall Assumption~\ref{asm:eq}). \red
\end{enumerate}
\end{secprob}

This problem can be solved by the following procedure. Define
\begin{align}
\beta := \sqrt{1 -\gamma^{-2}} >0.
\label{beta}
\end{align}
First, given~$\varepsilon \ge 0$, we find a solution~$P_\infty \succ 0$ ($P_\infty \in \bR^{n\times n}$) to the following GD Riccati inequality:
\begin{align}
&P_\infty \partial f(x) + \partial^\top f(x) P_\infty - \beta^2 P_\infty B B^\top P_\infty \nonumber\\
&+ C^\top C + \varepsilon P_\infty \preceq 0, \; \forall x \in \bR^n.
\label{Ric_Hcon}
\end{align}
This corresponds to the~$H_\infty$-control algebraic Riccati equation in the linear case. Similarly, one can infer a counterpart of the~$H_\infty$-filter algebraic Riccati equation (HFARE). However, when an equality is relaxed to an inequality, we need a modification to solve Problem~\ref{prob:Hcon}; a similar modification can be found in the Hamilton-Jacobi approach~\cite{LD:94}. In fact, we use a lower bound on the left-hand side of~\eqref{Ric_Hcon}, namely the following~$R_\infty (x) \succeq 0$ ($R_\infty (x) \in \bR^{n\times n}$), $x \in \bR^n$:
\begin{align}
- R_\infty (x) \preceq & \; P_\infty \partial f(x) + \partial^\top f(x) P_\infty - \beta^2 P_\infty B B^\top P_\infty \nonumber\\
&+ C^\top C + \varepsilon P_\infty, \; \forall x \in \bR^n.
\label{def_R}
\end{align}
Then, we consider the following inequality for~$Q_\infty \succ 0$ ($Q_\infty \in \bR^{n\times n}$) as an extension of HFARE:
\begin{align}
&\partial f(x) Q_\infty + Q_\infty \partial^\top f(x) - \beta^2 Q_\infty C^\top C Q_\infty \nonumber\\
&+ B B^\top + \varepsilon Q_\infty  \preceq - \gamma^{-2} Q_\infty R_\infty (x) Q_\infty, \; \forall x \in \bR^n.
\label{Ric_Hob}
\end{align}

As the first result of this section, we show that if the above GD Riccati inequalities~\eqref{Ric_Hcon} and~\eqref{Ric_Hob} have solutions, then Problem~\ref{prob:Hcon} can be solved.
\begin{secthm}\label{thm:Hcon}
For the system~\eqref{gsys}, suppose that
\begin{enumerate}
\renewcommand{\theenumi}{\Roman{enumi}}
\item given~$\varepsilon \ge 0$, the GR Riccati inequality~\eqref{Ric_Hcon} admits a solution~$P_\infty \succ 0$; 
\item for~$R_\infty (x) \succeq 0$, $x \in \bR^n$ satisfying~\eqref{def_R}, the GR Riccati inequality~\eqref{Ric_Hob} admits a solution~$Q_\infty \succ 0$;
\item $\gamma^2 > \lambda_{\max} (P_\infty Q_\infty)$.
\end{enumerate}
Then, the following dynamic controller solves item 1) of Problem~\ref{prob:Hcon}:
\begin{align}
K: \left\{\begin{array}{l}
\dot x_c = f(x_c) - \beta^2 B B^\top P_\infty x_c \\
\hspace{10mm} -  ( Q_\infty ^{-1} -  \gamma^{-2} P_\infty)^{-1} C^\top (C x_c - y),\\
u = - B^\top P_\infty x_c.
\end{array}\right.
\label{gcon}
\end{align}
Moreover, if~$\varepsilon > 0$, this controller solves item 2) of Problem~\ref{prob:Hcon}.
\red
\end{secthm}

Next, we investigate the connection between the $L_2$-gain of the closed-loop system~$\gamma>0$ and~$\varepsilon>0$. It is possible to show that $\gamma$ can be made smaller by selecting a smaller $\varepsilon$.
\begin{secthm}\label{thm:small_gamma}
For the system~\eqref{gsys}, suppose that items I) -- III) of Theorem~\ref{thm:Hcon} hold for~$\varepsilon >0$. Next, consider~$\varepsilon_1, \varepsilon_2 > 0$ and $0< \alpha < \beta^2$ such that~$\varepsilon = \varepsilon_1 + 2 \varepsilon_2$ and
\begin{align}
\alpha P_\infty B B^\top P_\infty \preceq \varepsilon_2 P_\infty,  \label{alpha_P}\\
\alpha Q_\infty C^\top C Q_\infty \preceq \varepsilon_2 Q_\infty. \label{alpha_Q}
\end{align}
Then, $P_\infty$, $\bar R_\infty (x):=R_\infty (x) + \varepsilon_2 P_\infty$, and $Q_\infty$ satisfy~\eqref{Ric_Hcon}, \eqref{def_R}, and~\eqref{Ric_Hob} for~$\bar \gamma := \gamma/\sqrt{1 + \alpha \gamma^2}$, respectively.
\red
\end{secthm}

Theorem \ref{thm:small_gamma} implies that $\gamma$ can be made smaller by finding $\alpha>0$ and $\varepsilon_2>0$ under the constraints \eqref{alpha_P}, \eqref{alpha_Q}, $\alpha < \beta^2$, and $2\varepsilon_2 < \varepsilon$. Namely, the GD Riccati inequalities \eqref{Ric_Hcon} and \eqref{Ric_Hob} hold for $\gamma = \bar \gamma$ if such $\alpha$ and $\varepsilon_2$ exist. Finding $\alpha$ and $\varepsilon_2$ can be formulated as a convex problem by relaxing $>$ to $\ge$ with a small positive constant. Furthermore, the maximum $\alpha$ can be computed. A larger $\alpha$ gives a smaller $\bar \gamma$, and a small $\gamma$ is beneficial to reduce the dimension of a controller while guaranteeing closed-loop stability as will be clear in Theorem~\ref{thm:clstab_rconH}. Moreover, for the new $\gamma = \bar \gamma$, we can again solve the GD Riccati inequalities to obtain better solutions, e.g., for reducing the dimension of a stabilizing controller.

\subsection{Controller Reduction}
As for GD balancing/LQG balancing, if the GD Riccati inequalities~\eqref{Ric_Hcon} and~\eqref{Ric_Hob} have solutions~$P_\infty \succ 0$ and $Q_\infty \succ 0$, respectively, then there exist coordinates, named \emph{GD $H_\infty$-balanced coordinates}, in which
\begin{align}
P_\infty = Q_\infty = \Pi := {\rm diag}\{\pi_1,\dots,\pi_n \}, \label{Hbal}\\ 
\pi_1 \ge \cdots \ge \pi_n > 0, \nonumber 
\end{align}
where we use the same symbol~$\Pi$ as GD LQG balancing by abuse of notation. Let~$\pi_r > \pi_{r+1}$, and define~$\Pi_1 := {\rm diag}\{\pi_1,\dots,\pi_r\}$. Hereafter, suppose that the realization~\eqref{gsys} is GD $H_\infty$-balanced. Then, in a similar manner as GD balancing/GD LQG balancing, a reduced-order model of the generalized plant~\eqref{gsys} can be constructed as follows:
\begin{align}
G_r: \left\{\begin{array}{l}
\dot x_r =  f_1(x_r,0) + B_1 u + B_1 w_{ur},\\
z_{ur} = u,\\
z_{yr} = y_r = C_1 x_r,\\
y = C_1 x_r + w_{yr},
\end{array}\right.
\label{rgsys}
\end{align}
where~$w_r (t):= [w_{ur}^\top (t)\; w_{yr}^\top (t)]^\top \in \bR^{m+p}$ and~$z_r(t) := [z_{ur}^\top (t)\; z_{yr}^\top (t)]^\top \in \bR^{m+p}$. For the sake of later analysis, we use different symbols from~$w$ and~$z$.

Note that~$\Pi_1$ satisfies both GD Riccati inequalities~\eqref{Ric_Hcon} and~\eqref{Ric_Hob} for the reduced-order model~\eqref{rgsys}, where the $(1,1)$-th block diagonal element of~$R_\infty (x)$ in the GD $H_\infty$-balanced coordinates is considered. Moreover, $\gamma^2 > \lambda_{\max} (P_\infty Q_\infty) = \lambda_{\max} (\Pi^2)=\pi_1^2$ implies $\gamma^2 > \lambda_{\max} (\Pi_1^2)=\pi_1^2$. Therefore, if all conditions in Theorem~\ref{thm:Hcon} and~$\pi_r > \pi_{r+1}$ hold, the following reduced-order controller solves Problem~\ref{prob:Hcon} for the reduced-order model~\eqref{rgsys}:
\begin{align}
K_r: \left\{\begin{array}{l}
\dot x_{cr} = f_1(x_{cr},0) - \beta^2 B_1 B_1^\top \Pi_1 x_{cr} \\
\hspace{10mm} -  ( \Pi_1^{-1} -  \gamma^{-2} \Pi_1)^{-1} C_1^\top (C_1 x_{cr} - y),\\
u = - B_1^\top \Pi_1 x_{cr}.
\end{array}\right.
\label{rgcon}
\end{align}

If this reduced-order controller satisfies some additional conditions, the closed-loop system consisting of the full-order plant~\eqref{gsys} and reduced-order controller~\eqref{rgcon}, denoted by~$\Gamma (G,K_r)$, is exponentially stable. The following theorem is the main result of this section, and the next subsection is dedicated to proving this.
\begin{secthm}\label{thm:clstab_rconH}
For the system~\eqref{gsys}, suppose that 
\begin{enumerate}
\renewcommand{\theenumi}{\Roman{enumi}}
\item items I) -- III) of Theorem~\ref{thm:Hcon} hold;
\item $\pi_r > \pi_{r+1}$;
\item $f$ is an odd function;
\item $\rho_r (1+\gamma \beta^{-1}) < 1$, where
\begin{align}
\rho_r := 2 \sum_{i=r+1}^n \frac{\beta \pi_i}{\sqrt{1 + \beta^2\pi_i^2}}.
\label{rho}
\end{align}
\end{enumerate}
Then, the following holds.
\begin{enumerate}
\item the closed-loop system~$\Gamma (G,K_r)$ consisting of the full-order plant~\eqref{gsys} and reduced-order controller~\eqref{rgcon} satisfies
\begin{align}
\| z \| \le \beta^{-1} \frac{\beta \gamma + \rho_r (\gamma + \beta)}{\beta - \rho_r (\gamma + \beta)} \| w \|
\end{align}
for the initial states~$x(0)=0$ and~$x_r(0)=0$.
\end{enumerate}
Furthermore, if~$\varepsilon > 0$ and~$w = 0$, then
\begin{enumerate}
\setcounter{enumi}{1}
\item $\Gamma (G,K_r)$ is GES at the origin.
\red
\end{enumerate}
\end{secthm}

\begin{secrem}
Even in the linear case, there is no inequality relaxation of $H_\infty$-balancing. Therefore, Theorem~\ref{thm:clstab_rconH} is a new result even for linear systems.
\red
\end{secrem}

Item IV) of Theorem \ref{thm:clstab_rconH} can be useful for deciding the dimension $r$ of a reduced-order controller. Given $\gamma>0$, \eqref{beta} determines $\beta$. Then, $\pi_i$ is obtained by solving the GD Riccati inequalities \eqref{Ric_Hcon} and \eqref{Ric_Hob} and by computing GD $H_\infty$-balanced coordinates as in \eqref{Hbal}. Thus, $r$ is only the design parameter in item IV) of Theorem \ref{thm:clstab_rconH}. The smallest $r$ satisfying the condition is the lowest dimension of a reduced-order controller for which Theorem \ref{thm:clstab_rconH} guarantees the closed-loop stability.

Note that our methods are based on matrix inequalities~\eqref{Ric_Hcon} -- \eqref{Ric_Hob}. They can be reduced into linear matrix inequalities (LMIs). For instance, consider~\eqref{Ric_Hcon} and define~$\hat P_\infty := P_\infty^{-1}$. Multiplying~$P_\infty^{-1}$ from both sides leads to the following LMIs for~$\hat P_\infty$:
\begin{align}
\left\{\begin{array}{l}
\hat P_\infty \succ 0,\\
\begin{bmatrix}
\partial f(x) \hat P_\infty + \hat P_\infty \partial^\top f(x) - \beta^2 B B^\top + \varepsilon \hat P_\infty & \hat P_\infty C^\top \\
C \hat P_\infty & -I
\end{bmatrix} \preceq 0.
\end{array}\right.
\label{LMI:Ric_Hcon}
\end{align}
This consists of an infinite family of LMIs for each fixed~$x\in\bR^n$. An approximation solution can be constructed for instance by taking finite sampling points $x_{(i)} \in \bR^n$, $i=1,\dots,D$. To find an exact solution, a convex relaxation can be used. Let~$\bA := \{A_1,\dots,A_L\} \subset \bR^{n \times n}$ be a family of matrices such that
\begin{align}
\partial f(x) \in {\rm ConvexHull} (\bA ), \; \forall x \in \bR^n.
\label{covexhull}
\end{align}
Then, any solution $\hat P_{\infty} \in \bR^{n\times n}$ to the following matrix inequalities
\begin{align}
\left\{\begin{array}{l}
\hat P_\infty \succ 0,\\
\begin{bmatrix}
A_i \hat P_\infty + \hat P_\infty A_i^\top - \beta^2 B B^\top + \varepsilon \hat P_\infty & \hat P_\infty C^\top \\
C \hat P_\infty & -I
\end{bmatrix} \preceq 0,\\
\hspace{60mm}i = 1,\dots,L.
\end{array}\right.
\label{LMI:Ric_Hcon2}
\end{align}
satisfies~\eqref{LMI:Ric_Hcon}. Indeed, from~\eqref{covexhull}, for each~$x \in \bR^n$, there exist $\theta_i(x) \ge 0$, $i=1,\dots,L$ such that $\partial f(x) = \sum_{i=1}^L \theta_i(x) A_i$ and $\sum_{i=1}^L \theta_i(x) = 1$. Then, multiplying $\theta_i(x)$ by the second LMI of~\eqref{LMI:Ric_Hcon2} and taking the summation for~$i=1,\dots,L$ lead to~\eqref{LMI:Ric_Hcon}. 

The set of LMIs~\eqref{LMI:Ric_Hcon2} can have multiple solutions~$\hat P_\infty$, but our results hold for all solutions. In controller reduction, $P_\infty$ having small eigenvalues is useful to guarantee the stability of the closed-loop system with a reduced-order controller. Since $\hat P_\infty$ corresponds to $P^{-1}_\infty$, a reasonable approach is to find $\hat P_\infty$ having the largest trace. This can be formulated as a convex optimization problem.

Similar statements hold for~\eqref{def_R} and~\eqref{Ric_Hob}. Therefore, our methods can be applied by solving finite families of LMIs. In other words, we do not need to solve nonlinear partial differential equations/inequalities differently from existing nonlinear controller reduction methods such as~\cite{PF:97,YW:01}. Depending on the system, $L$ can be large. However, the problem is still convex. In general, solving a convex (optimization) problem is more tractable than that of a nonlinear partial differential equation/inequality. We have similar remarks for GD balancing and GD LQG balancing. 

\begin{secrem}
A convex relaxation can be utilized for local model/controller reduction. As mention in Remark~\ref{rem:local}, under $f(0)=0$, results of GD balancing can be delivered to a convex subset, and this is also true for GD LQG balancing and GD $H_\infty$-balancing. Furthermore, the mentioned convex relaxation is always doable on a bounded set, since~$\partial f(x)$ is bounded on it. Therefore, we can always utilize the convex relaxation for model/controller reduction on a bounded convex set.
\red
\end{secrem}

The convex relaxation mentioned above is one approach. Even if $\partial f(x)$ is not bounded, a convex relaxation can be applied depending on problems. For instance, consider $\dot x = - x - x^3 + u$, where $\partial f(x) = -1-3x^2$ is not bounded. In this case, we can still find a solution to~\eqref{Lyap_con} by solving an LMI. The GD Lyapunov inequality~\eqref{Lyap_con} for $X>0$ becomes $2 X (-1-3x^2) + 1 + \varepsilon X \le 0$. Since $- X x^2 \le 0$ for all~$x \in \bR$, we only have to solve the following LMI: $- 2 X + 1 + \varepsilon X \le 0$. A solution is for instance $X=1$ for $\varepsilon =0.1$.

\subsection{Analysis for Proving Theorem~\ref{thm:clstab_rconH}}
\begin{figure}[t]
\begin{center}
\includegraphics[width=85mm]{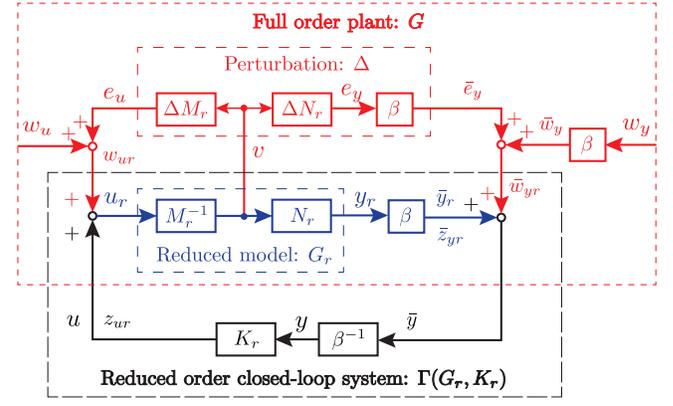}
\end{center}
\caption{The closed-loop system~$\Gamma (G,K_r)$ consisting of the full-order plant~\eqref{gsys} and reduced-order controller~\eqref{rgcon}}
\label{CLH:fig}
\end{figure}

The goal of this subsection is to prove Theorem~\ref{thm:clstab_rconH}. The main idea is to utilize the small gain theorem by interpreting the closed-loop system $\Gamma (G,K_r)$ as the feedback interconnection of the reduced-order closed-loop system (consisting of the reduced-order model~\eqref{gsys} and reduced-order controller~\eqref{rgcon}), denoted by~$\Gamma (G_r,K_r)$, and its perturbation~$\Delta$ as in Fig.~\ref{CLH:fig}.

Throughout this subsection, suppose that a realization~\eqref{gsys} is GD $H_\infty$-balanced, namely~\eqref{Hbal} holds. As a preliminary step, we summarize signals used in this subsection; also see Fig.~\ref{CLH:fig}.
\begin{align}
w_r &:= e + w, \; \bar w_r := \bar e + \bar w,\nonumber\\
\bar w_y &:= \beta w_y, \; \bar w_{yr} := \beta w_{yr}, \; \bar e_y := \beta e_y, \nonumber\\
e &:= \begin{bmatrix} e_u \\ e_y \end{bmatrix}, \; \bar e := \begin{bmatrix} e_u \\ \bar e_y \end{bmatrix}, \;
\bar w := \begin{bmatrix} w_u \\ \bar w_y \end{bmatrix}, \; \bar w_r := \begin{bmatrix} w_{ur} \\ \bar w_{yr} \end{bmatrix},\nonumber\\
u_r &:= u + w_{ur} = z_{ur} + w_{ur}, \nonumber\\
\bar y &:= \beta y, \; \bar y_r := \beta y_r, \nonumber\\
y&:=  y_r + w_{yr} = z_{yr} + w_{yr}.
\label{signals}
\end{align}

For the analysis of the feedback interconnection of~$\Delta$ and~$\Gamma (G_r,K_r)$, we estimate the gains of~$\Delta:v \to \bar e$ and~$\Gamma (G_r,K_r):\bar w + \bar e \to v$. Then, the small gain theorem provides a stability condition for the interconnection. However, the definitions of~$v$,~$e$, and~$\Delta$ are still missing. They are defined via a reduced-order model of the following system:
\begin{align}
\left\{\begin{array}{l}
\dot x = f (x) - \beta^2  B B^\top P_{\infty} x + B v, \\
\begin{bmatrix}
\bar z_y \\ u
\end{bmatrix}
=
\begin{bmatrix}
\beta C \\ - \beta^2 P_{\infty}
\end{bmatrix}x
+
\begin{bmatrix}
0 \\ I_m 
\end{bmatrix} v.
\end{array}\right.
\label{rcrHsys}
\end{align}
This is a GD RCR of the system~\eqref{sys} with the weighted output~$\bar z_y = \beta z_y = \beta Cx$ by~$\beta$. As Theorem~\ref{thm:rcrsys_Gram} for GD LQG balancing, its reduced-order model can be computed by using GD $H_\infty$-balancing of the system~\eqref{sys}; the proof is similar and thus is omitted.
\begin{secthm}\label{thm:rcrHsys_Gram}
Consider the system~\eqref{gsys}. Suppose that given~$\varepsilon \ge 0$, the GD Riccati inequalities~\eqref{Ric_Hcon} and~\eqref{Ric_Hob} admit solutions~$P_{\infty} \succ 0$ and~$Q_{\infty} \succ 0$, respectively. Then, $(\beta^2 P_\infty + Q_\infty^{-1})^{-1}$ (resp. $\beta^2 P_{\infty}$) is a GD controllability (resp. observability) Gramian of the GD RCR~\eqref{rcrHsys}.
\red
\end{secthm}

From Theorem~\ref{thm:rcrHsys_Gram}, in GD $H_\infty$-balanced coordinates of the system~\eqref{gsys}, the GD RCR~\eqref{rcrHsys} is also GD balanced (in the sense of Section~\ref{GD:sec}). Therefore, GD $H_\infty$-balanced truncation gives a GD balanced truncated reduced-order model of the GD RCR~\eqref{rcrHsys}:
\begin{align}
\left\{\begin{array}{l}
\dot x_r = f_1 (x_r,0) - \beta^2  B_1 B_1^\top \Pi_1 x_r + B_1 v, \\
\begin{bmatrix}
\bar z_{yr} \\ u_r
\end{bmatrix}
=
\begin{bmatrix}
\beta C_1 \\ - \beta^2 \Pi_1
\end{bmatrix}x_r
+
\begin{bmatrix}
0 \\ I_m 
\end{bmatrix} v.
\end{array}\right.
\label{rcrHrsys}
\end{align}
An important fact is that this is a GD RCR of the reduced-order model~\eqref{rgsys} with the weighted output~$\bar z_{yr} = \beta z_{yr} = \beta C_1 x_r$ by~$\beta$. 

Now, we are ready to introduce~$v$,~$e$, and~$\Delta$ in Fig.~\ref{CLH:fig}. First, we decompose the reduced-order GD RCR~\eqref{rcrHrsys} into two subsystems~$M_r^{-1}: u_r \to v$ and~$\beta N_r:v \to \bar z_{yr}$ found in Fig.~\ref{CLH:fig} as follows:
\begin{align}
M_r^{-1}&: \left\{\begin{array}{l}
\dot x_r = f_1(x_r, 0) + B_1 u_r, \\
v = \beta^2 B_1^\top \Pi_1 x_r + u_r,
\end{array}\right. \label{Mrsys}\\
\beta N_r&: \left\{\begin{array}{l}
\dot x_r = f_1(x_r, 0) - \beta^2 B_1 B_1^\top \Pi_1 x_r + B_1 v, \\
\bar z_{yr} = \beta z_{yr} = \beta C_1 x_r,
\end{array}\right.\nonumber
\end{align}
where the initial states~$x_r(0)$ of both subsystems are the same. One notices that their series interconnection~$\beta N_r \circ M_r^{-1}$ is the reduced-order model~\eqref{rsys} with the weighted output~$\bar z_{yr} = \beta z_{yr} = \beta C_1 x_r$ by~$\beta$, denoted by~$\beta G_r$ in Fig.~\ref{CLH:fig}.

Now,~$v$ is obtained as the output of~$M_r^{-1}$. To proceed with further analysis, we substitute
\begin{align*}
u_r = u + w_{ur} = - B_1^\top \Pi_1 x_{cr} + w_{ur}
\end{align*}
following from~\eqref{rgcon} and~\eqref{signals} into~$v$. Then,~$v$ is described as the output of the reduced-order closed-loop system~$\Gamma (G_r,K_r)$:
\begin{align*}
v = - \beta^2 B_1^\top \Pi_1 (x_{cr} - x_r) + w_{ur}.
\end{align*} 

According to the previous subsection, the gain from~$w_r$ to~$z_r$ of the reduced-order closed-loop system~$\Gamma (G_r,K_r)$ is not greater than~$\gamma$. Based on this, we can estimate an upper bound on the gain of~$\Gamma (G_r,K_r): \bar w_r \to v$ as stated below. For the sake of later analysis, we decompose~$\bar w_r$ as~$\bar w_r = \bar w + \bar e$, where~$\bar e$ will be introduced later.
\begin{seclem}\label{lem:Hcon_gain}
For the system~\eqref{gsys}, suppose that items I) and II) of Theorem~\ref{thm:clstab_rconH} hold. Then, for the reduced-order closed-loop system~$\Gamma (G_r, K_r)$, it holds that
\begin{align*}
\|v \| \le  (1+\gamma \beta^{-1}) \| \bar w_r \| \le  (1+\gamma \beta^{-1}) (\| \bar e \| + \| \bar w \|)
\end{align*}
for the initial states~$x_r(0)=0$ and~$x_{cr}(0)=0$. \red
\end{seclem}

Next, we estimate the gain of the perturbation~$\Delta$ in Fig.~\ref{CLH:fig} by using the GR RCR~\eqref{rcrHsys} and its reduced-order model~\eqref{rcrHrsys} again. One notices that~$\bar e_y = \bar z_y - \bar z_{yr}$ and~$e_u = z_u - z_{ur}$, i.e.
\begin{align*}
\Delta := \begin{bmatrix} \beta \Delta N_r & \Delta M_r \end{bmatrix}
= \begin{bmatrix} \beta N - \beta N_r & M - M_r \end{bmatrix}.
\end{align*}
Therefore,~$\Delta$ can be represented as error dynamics between the GD RCR~\eqref{rcrHsys} and its reduced-order model~\eqref{rcrHrsys}. Since model reduction is achieved by GD balanced truncation, an error bound is obtained as a direct application of Theorems~\ref{thm:error} and~\ref{thm:rcrHsys_Gram} without the proof.
\begin{seccor}\label{error_rcrHsys:cor}
For the system~\eqref{gsys}, suppose that items I) -- III) of Theorem~\ref{thm:clstab_rconH} hold. Then, for the GD RCR~\eqref{rcrHsys} and its reduced-order model~\eqref{rcrHrsys}, we have
\begin{align}
\| \bar e \| \le  \rho_r \| v \|, \; 
\rho_r := 2 \sum_{i=r+1}^n \frac{\beta \pi_i}{\sqrt{1 + \beta^2\pi_i^2}},
\label{error_rcrHsys}
\end{align}
where~$x(0)=0$ and~$x_r(0)=0$.
\red
\end{seccor}

Now, upper bounds on the gains of~$\Gamma (G_r,K_r): \bar w + \bar e \to v$ and~$\Delta: v \to \bar e$ are provided by Lemma~\ref{lem:Hcon_gain} and Corollary~\ref{error_rcrHsys:cor}, respectively. Therefore, the small gain theorem for their interconnection yields Theorem~\ref{thm:clstab_rconH}.

\section{Examples}\label{Ex:sec}
\subsection{Model Reduction by Generalized Differential Balancing}
In this subsection, model reduction by GD balanced truncation is illustrated by an example.
Consider the following system:
\begin{align}\label{sys:net}
\dot x_1 &= -3 x_1 + \sin (x_2 - x_1) + u, \nonumber\\
\dot x_i &= - 3 x_i + \sin (x_{i-1} - x_i) + \sin (x_{i+1} - x_i), \; i\neq 1,n,\nonumber\\
\dot x_n &= -3 x_n + \sin (x_{n-1} - x_n) ,\nonumber\\
y&= x_1.
\end{align}
Then, it follows that
\begin{align*}
&\partial f(x) \in {\rm ConvexHull} (\{A_0, A_1,\dots, A_{2n-2}\} ), \; \forall x \in \bR^n\\
&\hspace{4mm} A_0 := {\rm diag}\{-3,\dots,-3\}, \;
K = \begin{bmatrix}1 & -1 \\ -1 & 1 \end{bmatrix},\\
&\hspace{4mm} A_1 := A_0 + {\rm diag} \{K ,0,\dots,0 \}, \\
&\hspace{4mm} A_2 := A_0 + {\rm diag} \{-K ,0,\dots,0 \}, \\
&\hspace{4mm} A_3 := A_0 + {\rm diag} \{0, K ,0,\dots,0 \}, \\
&\hspace{4mm} A_4 := A_0 + {\rm diag} \{0, -K ,0,\dots,0 \}, \dots,\\
&\hspace{4mm} A_{2n-3} :=  A_0 + {\rm diag} \{0,\dots,0,K \},\\
&\hspace{4mm} A_{2n-2} :=  A_0 + {\rm diag} \{0,\dots,0,-K \}.
\end{align*}
Since each~$A_i$ is symmetric and~$B= C^\top$, both GD controllability and observability Gramians can be chosen to be the same, i.e. $X = Y$. 

\begin{figure}[tb]
\begin{center}
\includegraphics[width=85mm]{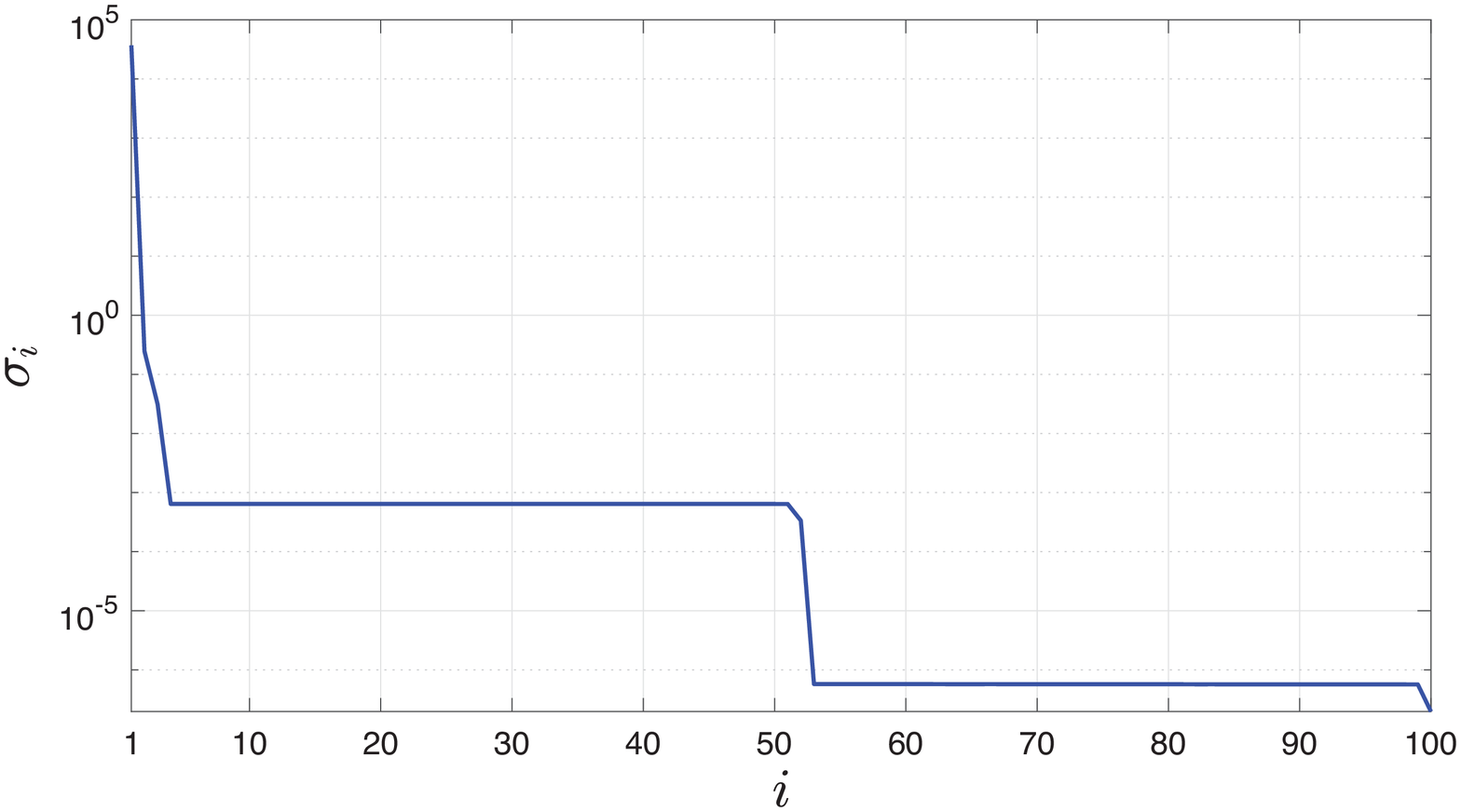}
\caption{Eigenvalues of a GD controllability/observability Gramian}
\label{eig:fig}
\vspace{6mm}
\includegraphics[width=85mm]{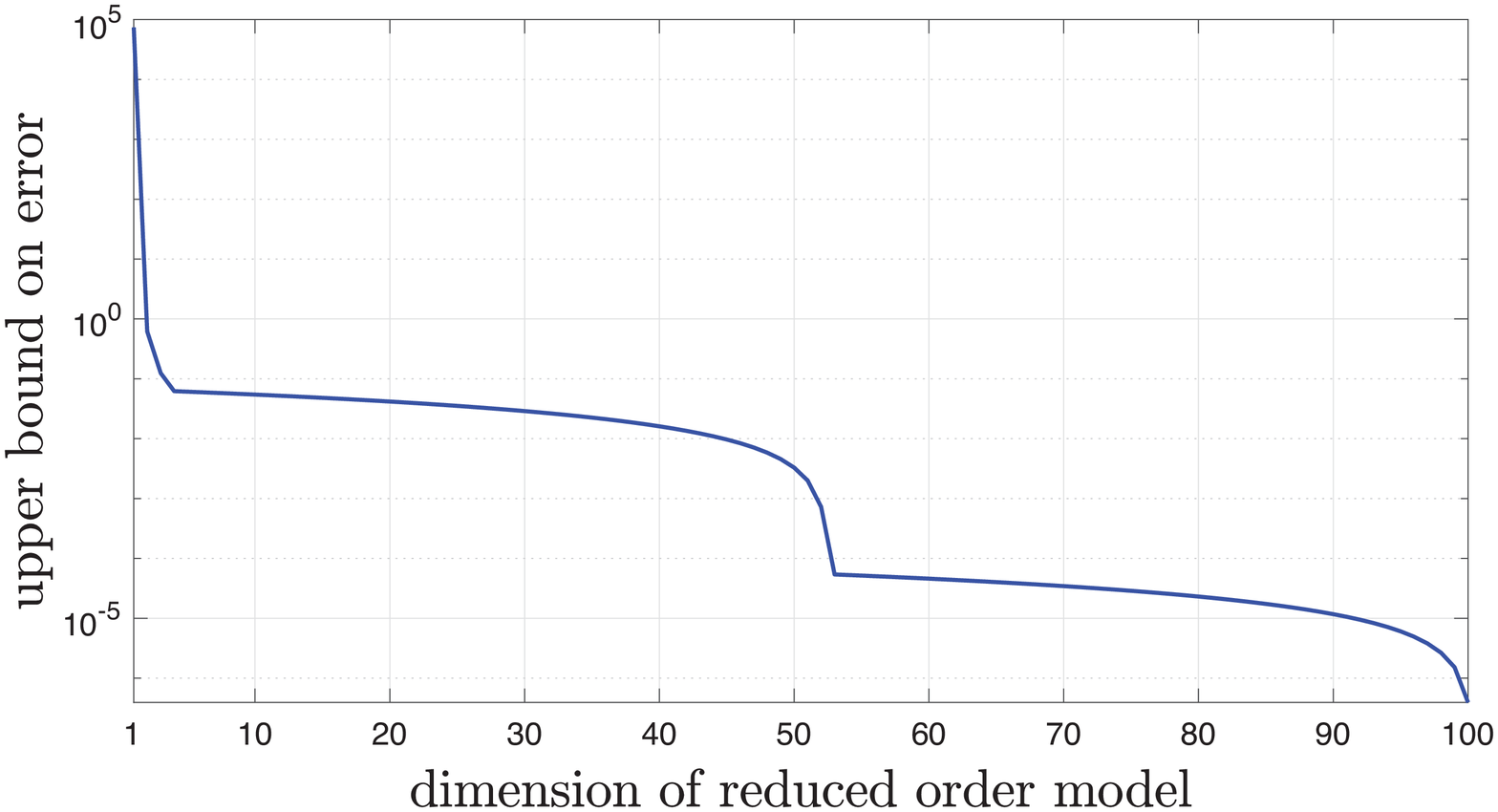}
\caption{Upper bounds on model reduction error by GD balanced truncation}
\label{error:fig}
\end{center}
\end{figure}

\begin{figure}[tb]
\begin{center}
\includegraphics[width=85mm]{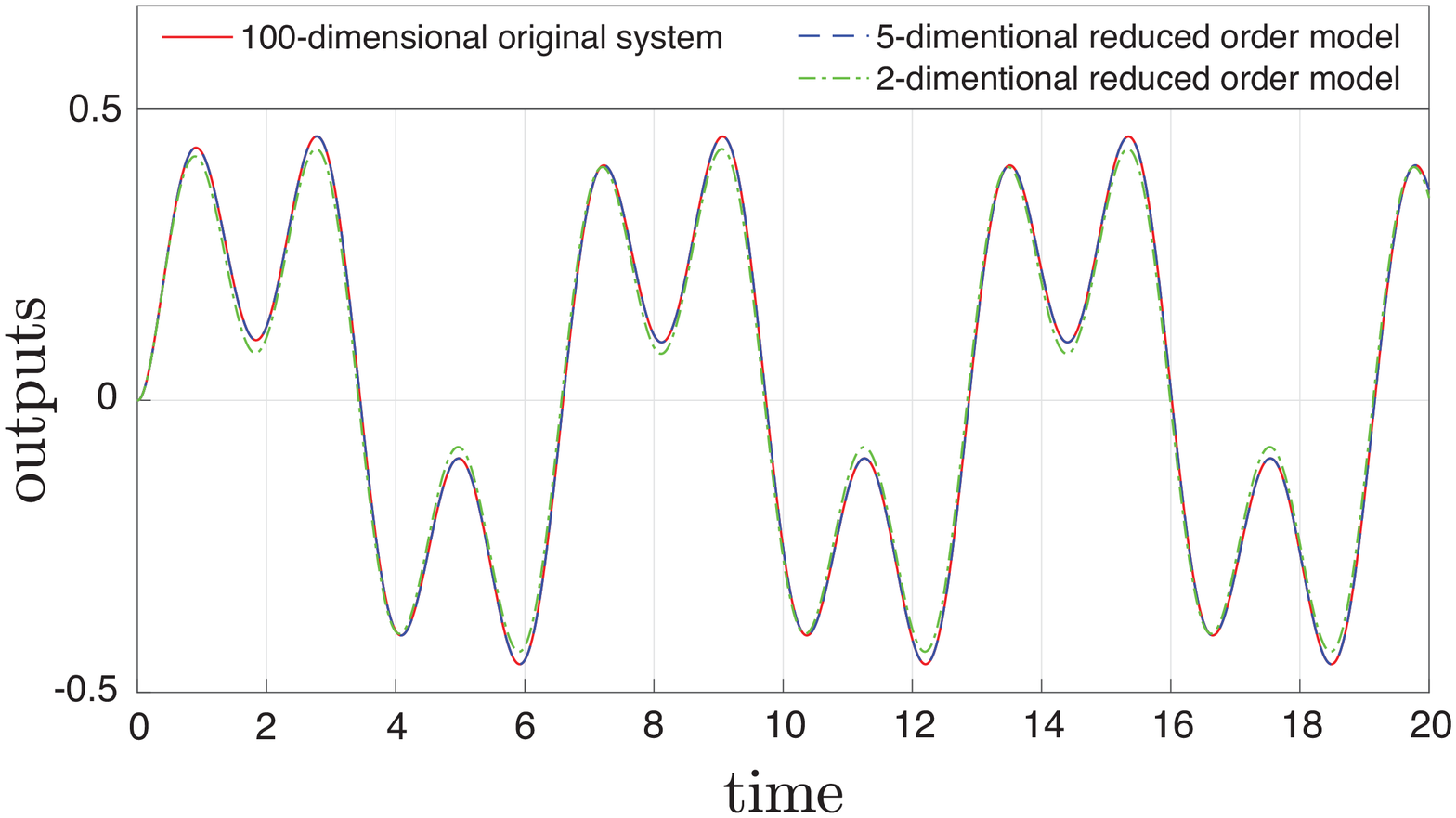}
\caption{Output trajectories of the system \eqref{sys:net} and its reduced order models when~$x(0)=0$,~$x_r(0)=0$, and~$u(t)=\sin (t) + \sin (3t)$}
\label{sim1:fig}
\vspace{7mm}
\includegraphics[width=80mm]{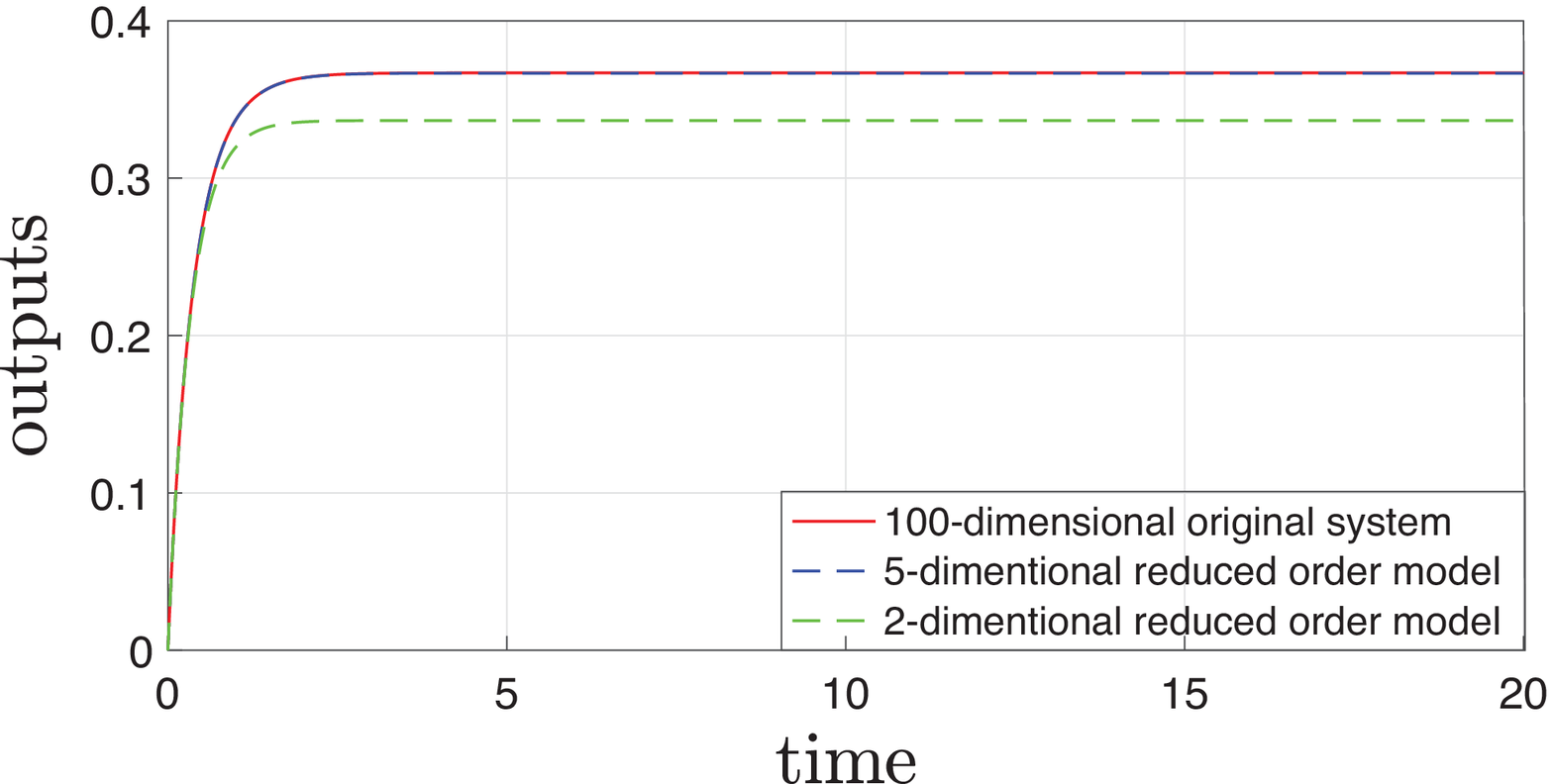}
\caption{Output trajectories of the system \eqref{sys:net} and its reduced order models when~$x(0)=0$,~$x_r(0)=0$, and~$u(t)=1$}
\label{sim2:fig}
\end{center}
\end{figure}

We consider the case where $n=100$ and solve the GD Lyapunov inequality \eqref{Lyap_con} with the convex relaxation.
Since it is difficult to display a solution $X$ due to its size, its eigenvalues are plotted in Fig.~\ref{eig:fig}.
Also, the upper error bounds~\eqref{error} for GD balanced truncation are plotted in Fig.~\ref{error:fig}. 
From the error bounds, it is expected that the error can be very large when the dimension of the reduced-order model is less than~$5$. To confirm this, we conduct simulations of the output trajectories of the $100$-dimensional original system as well as $5$-dimensional and $2$-dimensional reduced-order models. Fig.~\ref{sim1:fig} shows the output trajectories for~$x(0)=0$,~$x_r(0)=0$, and~$u(t)=\sin (t) + \sin (3t)$. Fig.~\ref{sim2:fig} shows the output trajectories for~$x(0)=0$,~$x_r(0)=0$, and~$u(t)=1$. As expected, the $2$-dimensional reduced-order model has a different behavior from the original system in contrast to the $5$-dimensional one. In other words, for this specific example, a $100$-dimensional nonlinear system can accurately be approximated by a $5$-dimensional reduced-order model by GD balancing.


\subsection{Controller Reduction by Generalized Differential $H_\infty$-Balancing}
In this subsection, controller reduction based on GD~$H_\infty$-balancing is illustrated by an example. Consider the following mechanical model controlled by a DC motor:
\begin{align}\label{mechanical_model}
\left\{\begin{array}{l}
\ddot z + c \dot z + k(z) = a i_c,\\
L \dot i_c + R i_c + k_v \dot z = u,\\
y = \begin{bmatrix}
z & \dot z
\end{bmatrix}^\top,
\end{array}\right.
\end{align}
where~$z$ and~$i_c$ denote the position of the mass and the current of the circuit, respectively. The control input~$u$ is the voltage. The parameters are~$c = 2$, $a = 1$, $L = 1/5$, $R = 1$, and~$k_v = 1$. The nonlinear function is chosen as~$k(x) = - \sin z$. Then,~$\partial k(z)/\partial z \in [-1, 1]$ for all~$z \in \bR$. 

We take the state as~$x := [z \; \dot z \; i_c]^\top$. To compute~$P_\infty$ and~$Q_\infty$, we use the convex relaxations like~\eqref{LMI:Ric_Hcon2} with
\begin{align*}
A_1 & =   \begin{bmatrix}
0 & 1 & 0\\
1 & -2 & 1\\
0 & -5 & -5
\end{bmatrix}, \;
A_2 =   \begin{bmatrix}
0 & 1 & 0\\
-1 & -2 & 1\\
0 & -5 & -5
\end{bmatrix}, \;
B = \begin{bmatrix}
0 \\ 0 \\ 5
\end{bmatrix},\\
C & =\begin{bmatrix}
1 & 0 & 0\\
0 & 1 & 0
\end{bmatrix},
\end{align*}
where~$A_1$ is not Hurwitz.

\begin{figure}[tb]
\begin{center}
\includegraphics[width=80mm]{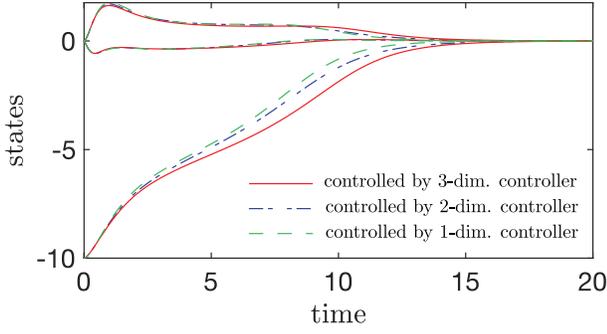}
\caption{State trajectories of the system \eqref{mechanical_model} controlled by controllers with different dimensions}
\label{sim:fig}
\end{center}
\end{figure}

Let~$\gamma^2 = 2$.  For~$\varepsilon = 0.01$,  solutions to~\eqref{LMI:Ric_Hcon2} and a similar relaxation of~\eqref{Ric_Hob} are obtained as
\begin{align*}
P_\infty &= \begin{bmatrix}
3.20 & 0.888 & 0.163\\
0.888 &  0.477 & 0.0778\\
0.163 & 0.0778 & 0.0154
\end{bmatrix},\\
Q_\infty &= \begin{bmatrix}
0.993 & -0.0848 &  0.104\\
-0.0848 & 0.338 & 0.0737\\
0.104 & 0.0737 &  2.43
\end{bmatrix}.
\end{align*}
Then, the matrix $\Pi$ in~\eqref{Hbal} is computed as follows:
\begin{align*}
\Pi = {\rm diag}\{ 1.78, 0.400, 0.192 \}.
\end{align*}
The conditions in Theorem~\ref{thm:clstab_rconH} hold for~$r = 2$, but not for~$r=1$. Therefore, from Theorem~\ref{thm:clstab_rconH}, the system can be stabilized by the full-order or $2$-dimensional reduced-order controller. Fig.~\ref{sim:fig} shows the trajectories of the system controlled by the full-order or reduced-order controllers. Although Theorem~\ref{thm:clstab_rconH} does not guarantee the stability for the $1$-dimensional reduced-order controller, all controllers stabilize the system in simulation. This implies that Theorem~\ref{thm:clstab_rconH} can be conservative, since this theorem is derived based on the small gain theorem. Developing a less conservative stability condition is included in future work.

\section{Conclusion}\label{Con:sec}
In this paper, we have developed GD balancing methods. First, for GD balancing, we have shown that a nonlinear system admitting a GD controllability Gramian is IES as a control system and have estimated an upper error bound for GD balanced truncation. Next, we have proposed GD LQG balancing that has been studied relating to GD RCR/LCR. As a byproduct, we have obtained observer-based stabilizing controllers while the separation principle does not work for nonlinear systems in general. Finally, we have established GD $H_\infty$-balancing for controller reduction and presented a stability condition for the closed-loop system consisting of the full-order  system and a reduced-order controller. These three GD balancing methods can be relaxed to LMIs. Future work includes extending the results of this paper to more general nonlinear systems by considering matrix-valued solutions to GD Lyapunov or Riccati inequalities.

\appendices
\section{Proofs for GD Balancing}
\subsection{Proof of Theorem~\ref{thm:Lyap_con}}
Before proving each item, we show that for each input~$u \in \cU$, the solution to the system \eqref{sys} exists.  Using~\eqref{Lyap_con}, compute
\begin{align*}
&\frac{d}{dt} |f(x) + B u|_{X^{-1}}^2 \\
&=(f(x) + B u)^\top X^{-1}\\
&\hspace{10mm}\times  (\partial f(x) X+ X \partial^\top f(x) ) X^{-1} (f(x) + B u)\\
&\hspace{4mm} + 2 \dot u^\top B^\top X^{-1} (f(x) + B u)\\
&\le - | B^\top  X^{-1} (f(x) + B u) |_2 - \varepsilon  |f(x) + B u|_{X^{-1}}^2\\
&\hspace{4mm} + 2\dot u^\top B^\top X^{-1} (f(x) + B u)\\
&= - \varepsilon  |f(x) + B u|_{X^{-1}}^2 - | B^\top  X^{-1} (f(x) + B u) - \dot u|^2 + |\dot u|^2\\
&\le - \varepsilon  |f(x) + B u|_{X^{-1}}^2 + |\dot u|^2.
\end{align*}
With the comparison principle~\cite[Lemma 3.4]{Khalil:02}, the time integration over the interval $[0,t]$ yields
\begin{align}
|f(x(t)) + B u(t)|_{X^{-1}}^2 \le & \; e^{- \varepsilon t} |f(x(0)) + B u(0)|_{X^{-1}}^2 \nonumber\\
&+ \int_0^t e^{- \varepsilon (t - \tau)} |\dot u(\tau)|^2 d\tau. 
\label{Lyap_con_bd0}
\end{align}
Therefore,~$|\dot x(t)|=|f(x(t)) + B u(t)|$ is a bounded function of~$t \in \bR_+$ for each~$u \in \cU$. This implies that~$|x(t)|$ exists for any~$t \in \bR_+$, but does not guarantee the boundedness of~$|x(t)|$. 

Now we are ready to prove each item.
\begin{IEEEproof}[Proof of item~1)]
Consider the line segment~$\gamma (s) = s x + (1-s) x'$,~$s \in [0,1]$. Then,
\begin{align}
f(x)- f(x') 
&= \int_0^1 \frac{d f(\gamma (s))}{d s}  ds \nonumber\\
& = \int_0^1 \partial f(\gamma (s)) (x - x') ds.
\label{path_int}
\end{align}
Using~\eqref{Lyap_con} and~\eqref{path_int}, similar computations as the previous give
\begin{align}
&\frac{d}{dt} |x - x'|_{X^{-1}}^2 \nonumber\\
&= 2(x - x')^\top X^{-1} (f(x)- f(x') + B (u - u')) \nonumber\\
&= \int_0^1  (x - x')^\top X^{-1} \nonumber\\
&\hspace{15mm} \times (\partial f(\gamma (s)) X + X \partial^\top f(\gamma (s))) X^{-1} (x -x') ds \nonumber\\
&\hspace{5mm}+ 2(x - x')^\top X^{-1} B (u - u') \nonumber\\
&\le - |B^\top X^{-1} (	x -x')|^2 - \varepsilon |x - x'|_{X^{-1}}^2 \nonumber\\
&\hspace{5mm} + 2(u - u')^\top B^\top X^{-1}  (x - x')\nonumber\\
&=  -\big|u - u' - B^\top X^{-1} (x -x') \big|^2 + | u - u' |^2 - \varepsilon |x - x'|_{X^{-1}}^2 \nonumber\\
&\le  - \varepsilon |x - x'|_{X^{-1}}^2 + | u - u' |^2 .
\label{eq:key2}
\end{align}
With the comparison principle, the time integration over the interval $[0,t]$ yields
\begin{align}
|x(t ) - x' (t )|_{X^{-1}}^2 \le & \; e^{- \varepsilon t} |x(0) - x'(0)|_{X^{-1}}^2 \nonumber\\
&+ \int_0^t e^{- \varepsilon (t - \tau)} | u(\tau) - u'(\tau)|^2 d\tau. 
\label{eq:key}
\end{align}
Recall~$x'(t) = \phi(t,x'_0,u') \in \bR^n$, $t \in \bR_+$ for each~$x'_0 \in \bR^n$ and $u' \in \cU$.
Thus,~\eqref{eq:key} implies the statement of item~1).
\end{IEEEproof}

\begin{IEEEproof}[Proof of item~2)]
The time integration of~\eqref{eq:key2} over the interval $[0, t]$ yields
\begin{align}
&|x(t) - x' (t)|_{X^{-1}}^2  + \varepsilon \int_0^t |x(\tau ) - x' (\tau )|_{X^{-1}}^2 d\tau \nonumber\\
&\le |x_0 - x'_0|_{X^{-1}}^2 + \int_0^t  | u(\tau ) - u' (\tau ) |^2  d\tau , \; \forall \tau > 0. 
\label{Lyap_con_bd2}
\end{align}
Since~$\varepsilon > 0$ and $u - u' \in L_2^m[0, \infty)$, taking~$\tau \to \infty$ implies~$x-x' \in L_2^n[0,\infty)$. Note that any~$x(t)$ with $u \in \cU$ is uniformly continuous with respect to~$t \in \bR_+$, since its derivative~$\dot x(t) = f(x(t))+ Bu(t)$ is bounded as shown in~\eqref{Lyap_con_bd0}. Therefore, the statement of item~2) follows from Barbalat's lemma~\cite[Lemma~8.2]{Khalil:02}. 
\end{IEEEproof}

\begin{IEEEproof}[Proof of item~3)] 
IES follows from~\eqref{eq:key} with~$u=u'$.
\end{IEEEproof}

\begin{IEEEproof}[Proof of item~4)]
For~$u = u' = u^* \in \bR^m$, \eqref{eq:key} implies that
\begin{align}
&|\phi (t, x(0), u^* ) - \phi (t, x'(0), u^* )|_{X^{-1}}^2\nonumber\\
& \le e^{- \varepsilon t} |x(0) - x'(0)|_{X^{-1}}^2, \; \forall t \ge 0, 
\label{inq:conmap}
\end{align}
for any $(x(0), x'(0)) \in \bR^n \times \bR^n$.
Fix some~$\bar t> 0$. Then, $e^{- \varepsilon \bar t} < 1$, and thus~$x \mapsto \phi (\bar t, x, u^* )$ is a contraction mapping. From the Banach fixed point theorem~\cite{Banach:22}, there exists a unique~$x^* \in \bR^n$ such that~$\lim_{n \to \infty} \phi (n \bar t, \hat x, u^* ) = x^*$ as~$n=1,2,\dots$ for any~$\hat x \in \bR^n$. Substituting~$x(0) = x^*$ and~$x'(0) = \phi (n \bar t, \hat x, u^* )$ into~\eqref{inq:conmap} yields
\begin{align*}
&|\phi (t, x^*, u^* ) - \phi (n \bar t , \phi(t, \hat x, u^*), u^* )|_{X^{-1}}^2\\
&=|\phi (t, x^*, u^* ) - \phi (t , \phi(n \bar t, \hat x, u^*), u^* )|_{X^{-1}}^2\\
& \le e^{- \varepsilon t} |x^* - \phi (n \bar t, \hat x, u^* )|_{X^{-1}}^2, \; \forall t \ge 0.
\end{align*}
Taking~$n \to \infty$ as~$n=1,2,\dots$ leads to
\begin{align*}
|\phi (t, x^*, u^* ) - x^*|_{X^{-1}}^2 = 0, \; \forall t \ge 0.
\end{align*}
Thus, $x^*$ is an equilibrium point of the system and GES.
\end{IEEEproof}

\begin{IEEEproof}[Proof of item~5)]
This can be confirmed by substituting~$u'=0$ and the corresponding equilibrium point~$x'=x^*$ into~\eqref{eq:key}.
\end{IEEEproof}

\subsection{Proof of Theorem~\ref{thm:Lyap_ob}}
\begin{IEEEproof}
Similarly to the proof of Theorem~\ref{thm:Lyap_con}, one can compute the following with~\eqref{Lyap_ob},
\begin{align*}
\frac{d}{dt} |f(x) + B u|_Y^2 
&\leq  - \varepsilon |f(x) + B u|_{Y}^2 - |C ( f(x) + B u )|^2\\
&\hspace{5mm} + 2\dot u^\top B^\top Y (f(x) + B u).
\end{align*}
and
\begin{align}
\frac{d}{dt} |x - x'|_Y^2
&\leq - \varepsilon |x - x'|_{Y}^2 - |y -y'|^2 \nonumber\\
&\hspace{5mm} + 2(u - u')^\top B^\top Y (x - x'). \label{key:Lyap_ob}
\end{align}
For~$u=u'=u^* \in \bR^m$, one has the statements of the theorem as the proof of Theorem~\ref{thm:Lyap_con}.
\end{IEEEproof}

\subsection{Proof of Corollary~\ref{cor:mapping}}
\begin{IEEEproof}
According to~\cite[Corollary 4.3]{Palais:59}, it suffices to show that~$\partial f(x)$ is non-singular at each~$x \in \bR^n$, and~$\lim_{|x| \to \infty} |f(x)| \to \infty$. From~\eqref{Lyap_con}, one notices that all real parts of the eigenvalues of~$\partial f(x)$ is not greater than~$-\varepsilon /2 < 0$, which implies that~$\partial f(x)$ is non-singular at each~$x \in \bR^n$. 

Next, from item~4) of Theorem~\ref{thm:Lyap_con}, there exists~$x^* \in \bR^n$ such that~$f(x^*) = 0$. Substituting~$x' = x^*$ and~$u=u'=0$ into~\eqref{eq:key2} leads to
\begin{align*}
 |x - x^*|_{X^{-1}}^2 + (2/\varepsilon) (x - x^*)^\top X^{-1} f(x) \le 0.
\end{align*}
Adding $|f(x) /\varepsilon|_{X^{-1}}^2$ to both sides yields
\begin{align*}
 |x - x^* + f(x)/\varepsilon|_{X^{-1}}  \le  |f(x) /\varepsilon|_{X^{-1}},
\end{align*}
From the triangular inequality, we have
\begin{align*}
 |x - x^*|_{X^{-1}} - | f(x)/\varepsilon|_{X^{-1}}  \le  |f(x) /\varepsilon|_{X^{-1}}.
\end{align*}
Therefore, $|f(x)| \to \infty$ as $|x| \to \infty$.
\end{IEEEproof}

\subsection{Proof of Theorem~\ref{thm:con_lowbd}}
\begin{IEEEproof}
In~\eqref{eq:key2}, taking the time integration in~$[-t,0]$ instead of~$[0,t]$ yields
\begin{align*}
|x_0 - x'_0|_{X^{-1}}^2 &\le |x(-t) - x'(-t)|_{X^{-1}}^2\\
&\hspace{5mm} + \int_{-t}^0  | u(\tau) - u' (\tau) |^2  d\tau , \; \forall t > 0. 
\end{align*}
Therefore,~\eqref{con_func_lb} holds for any pair~$(u,u')$ achieving~$\lim_{t \to \infty}(x(-t) - x'(-t)) =0$.
\end{IEEEproof}

\subsection{Proof of Theorem~\ref{thm:ob_upbd}}
\begin{IEEEproof}
The time integration in $[0,t]$ of~\eqref{key:Lyap_ob} with~$u = u'=u^*$ leads to
\begin{align*}
|x(t) - x'(t)|_Y^2 + \int_0^t  | y(\tau) - y' (\tau) |^2  d\tau
\le |x_0 - x'_0|_Y^2, \; \forall t > 0. 
\end{align*}
Since the system has the detectability property~\eqref{detectability}, we have~\eqref{ob_func_ub} by taking~$t \to \infty$; see Remark~\ref{rem:epsilon} also.
\end{IEEEproof}

\subsection{Proof of Theorem~\ref{thm:error}}
\begin{IEEEproof}
Suppose that~$r = n-1$. Define a function~$V$ as
\begin{align}
V(x,x_r) = \sigma_n^{-2}
\left| x - c - \begin{bmatrix}  x_r \\ 0 \end{bmatrix} \right|_\Sigma^2  
+\left| x -c - \begin{bmatrix}  -x_r \\ 0 \end{bmatrix} \right|_{\Sigma^{-1}}^2.
\label{def:Verror}
\end{align}
Its time derivative along the trajectories of the systems~\eqref{sys} and~\eqref{rsys} can be shown to read
\begin{align*}
\frac{dV(x,x_r)}{dt}
= \sigma_n^{-2} W_1(x,x_r) + W_2(x,x_r) + W_3(x,x_r),
\end{align*}
where
\begin{align*}
W_1(x,x_r) := \; &2\left( x - c -\begin{bmatrix}  x_r \\ 0 \end{bmatrix} \right)^{\!\top}\\
&\times \Sigma  
\left( f(x) - \begin{bmatrix}  f_1(x_r + c_1,c_2) \\ 0 \end{bmatrix} \right),\\
W_2(x,x_r) := \; &2\left( x - c - \begin{bmatrix}  -x_r \\ 0 \end{bmatrix} \right)^{\!\top}\\
&\times \Sigma^{-1}\!
\left( f(x) + \begin{bmatrix}  f_1(x_r + c_1,c_2) \\ 0 \end{bmatrix} \right),\\
W_3(x,x_r) := \; &4 \left( x - c - \begin{bmatrix}  - x_r \\ 0 \end{bmatrix} \right)^{\!\top} 
\Sigma^{-1} B u.
\end{align*}

In order to proceed with further computations, define
\begin{align*}
\gamma_-(s)  &= s x + (1 -s) \left( c + \begin{bmatrix}  x_r \\ 0 \end{bmatrix} \right) ,\\
\gamma_+(s)  &= s x + (1 -s) \left( c + \begin{bmatrix}  -x_r \\ 0 \end{bmatrix} \right).
\end{align*}
Then,
\begin{align}
f(x) -  f(x_r+ c_1,c_2) 
= \int_0^1  \partial f(\gamma_-(s)) ds  \left( x - c - \begin{bmatrix}  x_r \\ 0 \end{bmatrix} \right).
\label{path1}
\end{align}
Also, from~$f(x_r + c_1,c_2)= -f(-x_r+ c_1,c_2)$,
\begin{align}
&f(x) + f(x_r+ c_1,c_2) \nonumber\\
&= f(x) - f(-x_r+ c_1,c_2)  \nonumber\\
&= \int_0^1 \partial f(\gamma_+(s)) ds  \left( x - c - \begin{bmatrix}  -x_r \\ 0 \end{bmatrix} \right).
\label{path2}
\end{align}

It follows from~\eqref{Lyap_ob} and~\eqref{path1} that
\begin{align*}
&W_1(x,x_r)\\
&= 2\left( x - c - \begin{bmatrix}  x_r \\ 0 \end{bmatrix} \right)^{\!\top}
\Sigma \big( f(x) - f(x_r + c_1,c_2) \big)\\
&\hspace{5mm}+ 2\left( x - c - \begin{bmatrix}  x_r \\ 0 \end{bmatrix} \right)^{\!\top}
\Sigma  
\begin{bmatrix}  0 \\ f_2(x_r + c_1,c_2) \end{bmatrix} \\
&=\left( x - c - \begin{bmatrix}  x_r \\ 0 \end{bmatrix} \right)^{\!\top}
\int_0^1 \Big(\Sigma \partial f(\gamma_-(s)) \\
&\hspace{17mm}+ \partial^\top f(\gamma_-(s)) \Sigma \Big) ds  \left( x - c - \begin{bmatrix}  x_r \\ 0 \end{bmatrix} \right)\\
&\hspace{5mm} + 2\sigma_n ( x_2 - c_2 )^\top f_2(x_r + c_1,c_2)\\ 
&\le - | y - y_r |^2 + 2\sigma_n  ( x_2 - c_2 )^\top f_2(x_r + c_1,c_2)\\
&\hspace{5mm} - \varepsilon \left| x - c - \begin{bmatrix}  x_r \\ 0 \end{bmatrix} \right|_{\Sigma}^2.
\end{align*}

Similarly, it follows from~\eqref{Lyap_con} and~\eqref{path2} that
\begin{align*}
&W_2(x,x_r)\\
&= 2\left( x - c - \begin{bmatrix}  -x_r \\ 0 \end{bmatrix} \right)^{\!\top}
\Sigma^{-1} 
\big( f(x) + f(x_r + c_1,c_2)\big)\\
&\hspace{5mm}-2\left( x - c  - \begin{bmatrix}  -x_r \\ 0 \end{bmatrix} \right)^{\!\top}
\Sigma^{-1} 
\begin{bmatrix}  0 \\ f_2(x_r + c_1,c_2) \end{bmatrix} \\
&= \left( x - c - \begin{bmatrix}  -x_r \\ 0 \end{bmatrix} \right)^{\!\top}
\int_0^1 \Big(\Sigma^{-1} \partial f(\gamma_+(s))\\
&\hspace{15mm} + \partial f(\gamma_+(s)) \Sigma^{-1} \Big) ds  \left( x - c - \begin{bmatrix}  -x_r \\ 0 \end{bmatrix} \right)\\
&\hspace{5mm} - 2\sigma_n^{-1} (x_2 - c_2)^\top f_2(x_r+c_1,c_2)\\
&\le -\left|  B^\top \Sigma^{-1}\left( x - c - \begin{bmatrix}  -x_r \\ 0 \end{bmatrix} \right)\right|^2\\
&\hspace{5mm}- 2\sigma_n^{-1} (x_2 - c_2)^\top f_2(x_r+c_1,c_2) \\
&\hspace{5mm}- \varepsilon \left| x - c - \begin{bmatrix} - x_r \\ 0 \end{bmatrix} \right|_{\Sigma^{-1}}^2.
\end{align*}

Combining the above computations leads to
\begin{align}
\frac{dV(x,x_r)}{dt}
&\le - \sigma_n^{-2} | y - y_r |^2 + 4 | u|^2 - \varepsilon V(x,x_r) \nonumber\\
&\hspace{5mm} - \left| 2u - B^\top \Sigma^{-1}\left( x - c - \begin{bmatrix}  -x_r \\ 0 \end{bmatrix}  \right)\right|^2 \nonumber\\
&\le - \sigma_n^{-2}  | y - y_r |^2 + 4 | u|^2 - \varepsilon V(x,x_r). \nonumber
\end{align}
The time integration over the interval $[0,t]$ yields
\begin{align}
&\sigma_n^2 V(x(t ),x_r(t )) \nonumber\\
&\hspace{5mm} + \varepsilon \int_0^t \sigma_n^2 V(x(\tau ),x_r(\tau )) d\tau +  \int_0^t | y(\tau ) - y_r(\tau ) |^2 d\tau \nonumber\\
&\le \sigma_n^2 V(x(0),x_r(0)) + 4 \sigma_n^2 \int_0^t | u(\tau) |^2 d\tau, \; \forall t > 0. \label{error_int}
\end{align}
Note that $V(c,0) = 0$ and $V(x,x_r) \ge 0$ for all $(x,x_r)\in\bR^n\times\bR^r$. Then, for~$x(0)=c$ and~$x_r(0)=0$, we obtain
\begin{align*}
\| y - y_r\| \le 2 \sigma_n \|u\|.
\end{align*}
Recursively using the triangular inequality leads to~\eqref{error}. Although~$\sigma_{n-1}>\sigma_n$ is assumed, the proof can be extended to the case where~$\sigma_r>\sigma_{r+1} \ge \cdots \ge \sigma_n$. 
\end{IEEEproof}

\subsection{Preliminary Analysis for Proving Theorem~\ref{thm:clstab_rconH}}
For the sake of latter analysis, we compute a dissipation type inequality for two-dimensional model reduction obtained from~\eqref{error_int} when~$c=0$. From the definition of~$V(x,x_r)$ in~\eqref{def:Verror} with~$c=0$, there exists~$\underline{c}_n>0$ such that~$\underline{c}_n |x|^2 \le  \sigma_n^2 V(x,x_r)$. Also, when $x_r = 0$, there exists~$\overline{c}_n>0$ such that~$\sigma_n^2 V(x,0) \le \overline{c}_n |x|^2$. Therefore,~\eqref{error_int} with~$x_r(0)=0$ leads to 
\begin{align*}
&\underline{c}_n |x(t )|^2 + \varepsilon \int_0^t \underline{c}_n |x(\tau )|^2 d\tau +  \int_0^t | y(\tau ) - y_r(\tau ) |^2 d\tau \nonumber\\
&\le \overline{c}_n |x(0)|^2 + 4 \sigma_n^2 \int_0^t | u(\tau) |^2 d\tau, \; \forall t > 0.
\end{align*}

Next, let~$\bar x_r$ and~$\bar y_r$ be the state and output of the~$n-2$-dimensional reduced-order model, and~$\bar V(x_r, \bar x_r)$ denote the corresponding function to~\eqref{def:Verror} with~$c = 0$. Again, there exist~$\underline{c}_{n-1}>0$ such that~$\underline{c}_{n-1} |\bar x_r|^2 \le  \sigma_{n-1}^2 V(x_r,\bar x_r)$ and~$\overline{c}_{n-1}>0$ such that for $x_r =0$,~$\sigma_n^2 V(0,\bar x_r) \le \overline{c}_{n-1} |\bar x_r|^2$. Again,~\eqref{error_int} with~$x_r(0)=0$ leads to  
\begin{align*}
&\underline{c}_{n-1} |\bar x_r (t )|^2 + \varepsilon \int_0^t \underline{c}_{n-1} |\bar x_r (\tau )|^2 d\tau +  \int_0^t | y_r(\tau ) - \bar y_r(\tau ) |^2 d\tau \nonumber\\
&\le \overline{c}_{n-1} |\bar x_r(0)|^2 + 4 \sigma_{n-1}^2 \int_0^t | u(\tau) |^2 d\tau, \; \forall t > 0.
\end{align*}

Combining these two yields
\begin{align*}
&\left( \underline{c} ( |x| + |\bar x_r|)^2 + \varepsilon \int_0^t \underline{c} ( |x| + |\bar x_r| )^2 d\tau +  \int_0^t | y - \bar y_r |^2 d\tau  \right)^{1/2}\\
&\le \left( \underline{c}_n |x|^2 + \varepsilon \int_0^t \underline{c}_n |x|^2 d\tau +  \int_0^t | y - y_r |^2 d\tau  \right)^{1/2}\\
&\hspace{4mm} +\left( \underline{c}_{n-1} |\bar x_r|^2 + \varepsilon \int_0^t \underline{c}_{n-1} |\bar x_r |^2 d\tau +  \int_0^t | y_r - \bar y_r |^2 d\tau \right)^{1/2}\\
&\le \left( \overline{c}_n |x(0)|^2 + 4 \sigma_n^2 \int_0^t | u(\tau) |^2 d\tau \right)^{1/2}\\
&\hspace{4mm} + \left( \overline{c}_{n-1} |\bar x_r(0)|^2 + 4 \sigma_{n-1}^2 \int_0^t | u(\tau) |^2 d\tau \right)^{1/2}\\
&\le \overline{c} ( |x(0)| + |\bar x_r(0)| ) + 2 (\sigma_n + \sigma_{n-1}) \| u \|
\end{align*}
for some constants~$\underline{c}$, $\overline{c}>0$, where the triangular inequality is used in the first and last inequalities. Equivalently, we have
\begin{align*}
&\underline{c} ( |x(t)| + |\bar x_r(t)| )^2 + \varepsilon \int_0^t \underline{c} ( |x| + |\bar x_r| )^2 d\tau +  \int_0^t | y - \bar y_r |^2 d\tau\\
&\le (\overline{c} ( |x(0)| + |\bar x_r(0)|) + 2 (\sigma_n + \sigma_{n-1}) \| u \|)^2.
\end{align*}
Note that, for any~$a, b \in \bR$ and~$\theta >0$, we have~$2ab \le (1/\theta)a^2 + \theta b^2$. Applying this twice to the right-hand side leads to the following for some~$\hat c > 0$,
\begin{align*}
&\underline{c} ( |x(t)| + |\bar x_r(t)| )^2 + \varepsilon \int_0^t \underline{c} ( |x| + |\bar x_r| )^2 d\tau +  \int_0^t | y - \bar y_r |^2 d\tau\\
&\le (1 + 1/\theta ) \hat{c}( |x(0)|^2 + |\bar x_r(0)|^2) \\
&\hspace{4mm}+ 4 (1+ \theta )(\sigma_{n-1} + \sigma_n)^2 \int_0^t | u(\tau) |^2 d\tau.
\end{align*} 

By repeating similar procedure, for the~$r$-dimensional reduced-order model, we have
\begin{align}
&\underline{c} ( |x(t)|^2 + |x_r(t)|)^2 + \varepsilon \int_0^t \underline{c}  ( |x(\tau)|^2 + |x_r(\tau)|)^2 d\tau \nonumber\\
&+  \int_0^t | y(\tau) - y_r(\tau) |^2 d\tau \nonumber\\
&\le  (1 + 1/\theta ) \overline{c}( |x(0)|^2 + |\bar x_r(0)|^2) \nonumber\\
&\hspace{4mm}+ 4 (1+ \theta ) \left(\sum_{i=r+1}^n \sigma_i \right)^2 \int_0^t | u(\tau) |^2 d\tau, \; \forall t \in \bR_+
\label{pre:clstab_rconH}
\end{align}
for any~$\theta>0$, where by abuse of notation, $x_r$ and $y_r$ denote the state and output of the~$r$-dimensional reduced-order model, and the initial states of the~$n-1$-to~$r+1$-reduced-order models are all zero.

\section{Proofs for GD LQG Balancing}
\subsection{Proof of Theorem~\ref{thm:up_post}}
\begin{IEEEproof}
Since $J_+$ is defined by taking the infimum, it suffices to show
\begin{align*}
\int_0^\infty  ( | u(\tau ) - u' (\tau ) |^2 + |y (\tau ) - y' (\tau ) |^2 )  d\tau  \le  |x_0 - x'_0|_P^2.
\end{align*}
for some pair~$(u,u')$ satisfying~$u - u' \in L_2^m [0, \infty)$ and achieving~$x(\infty) = x'(\infty)$ for any~$(x_0, x'_0) \in \bR^n \times \bR^n$. As such a pair of inputs, we choose~$u = - B^\top P x$ and~$u' = - B^\top P x'$. One notices that $P^{-1}$ is a GD controllability Gramian of the closed-loop system~$\dot x = f(x) - B B^\top P x + B v$. Applying~\eqref{Lyap_con_bd0} to the closed-loop system with~$v=0$ implies the existence of its solution for any~$t \ge 0$ and initial state~$x(0) \in \bR^n$ even when~$\varepsilon = 0$. In addition,~\eqref{eq:key} with~$x'(0) = 0$ and~$v = v' = 0$ concludes that the solution is a bounded function of~$t\ge 0$, and consequently uniformly continuous, where we use  Assumption~\ref{asm:eq} that the origin is an equilibrium point. 

In a similar manner as the proof of Theorem~\ref{thm:Lyap_con}, it follows from~\eqref{Ric_con} and~\eqref{path_int} that
\begin{align*}
&\frac{d}{dt} |x - x'|_P^2\\
&=2 (x - x')^\top P (f(x)- f(x') + B (u - u')) \\
&= \int_0^1 (x - x')^\top ( P \partial f(\gamma (s)) + \partial^\top f(\gamma (s)) P ) (x -x') ds\\
&\hspace{5mm}+2 (x - x')^\top P B (u - u') \\
&\le (x - x')^\top P B B^\top P (x -x')  \\
&\hspace{5mm}+ 2(x - x')^\top P B (u - u') - | y - y' |^2\\
&= |u - u' + B^\top P (x -x') |^2 - | u - u' |^2 - | y - y' |^2.
\end{align*}
The time integration over~$[0,t]$ yields
\begin{align*}
&|x(t ) - x' (t )|_P^2 - |x_0 - x'_0|_P^2 \\
& \le \int_0^t  | u(\tau ) - u' (\tau ) + B^\top P (x(\tau ) -x'(\tau )) |^2  dt \\
&\hspace{4mm} - \int_0^t  ( | u(\tau ) - u' (\tau ) |^2 + |y (\tau ) - y' (\tau ) |^2 )  d\tau .
\end{align*}
For~$u = - B^\top P x$ and~$u' = - B^\top P x'$, we have
\begin{align*}
\int_0^t  ( | u(\tau ) - u' (\tau ) |^2 + |y (\tau ) - y' (\tau ) |^2 )  d\tau  \le  |x_0 - x'_0|_P^2.
\end{align*}
Taking~$t \to \infty$ concludes~$u- u' = - B^\top P (x -x') \in L_2^m[0,\infty )$ for any pair~$(x_0,x_0') \in \bR^n \times \bR^n$. Since as mentioned above, $x(t)$ and~$x'(t)$ are uniformly continuous, Barbalat's lemma concludes
\begin{align*}
\lim_{t \to \infty} | u(t) - u'(t) |  = 0, \; \lim_{t \to \infty} |y(t) -y'(t) | =0.
\end{align*}
The detectability property~\eqref{detectability} implies~$\lim_{t \to \infty} | x(t) - x'(t) | = 0$.  
\end{IEEEproof}

\subsection{Proofs of Theorem~\ref{thm:low_past}}
\begin{IEEEproof}
In a similar manner as the proof of Theorem~\ref{thm:Lyap_con}, it follows from~\eqref{Ric_ob} that
\begin{align*}
&\frac{d}{dt} |x - x'|_{Q^{-1}}^2\\
&= 2(x - x')^\top Q^{-1} (f(x)- f(x') + B (u - u')) \\
&= \int_0^1 (x - x')^\top  Q^{-1}  ( \partial f(\gamma (s)) Q\\
&\hspace{35mm} +Q \partial^\top f(\gamma (s))) Q^{-1} (x -x') ds \\
&\hspace{5mm}+2 (x - x')^\top  Q^{-1}  B (u - u') \\
&\le  - (x - x')^\top Q^{-1}  B B^\top Q^{-1} (x -x')  \\
&\hspace{5mm}+ 2 (x -  x')^\top Q^{-1} B (u - u') + | y - y' |^2 \\
&\le - |u - u' - B^\top Q^{-1} (x -x') |^2 + | u - u' |^2 + | y - y' |^2\\
&\le | u - u' |^2 + | y - y' |^2.
\end{align*}
The time integration over~$[-t,0]$ yields
\begin{align*}
|x_0 - x'_0|_{Q^{-1}}^2 
\le&\; |x(-t ) - x' (-t )|_{Q^{-1}}^2 \\
& + \int_{-t}^0  ( | u(\tau ) - u' (\tau ) |^2 + |y (\tau ) - y' (\tau ) |^2 )  d\tau .
\end{align*}
This inequality holds for arbitrary pair~$(u,u')$. By restricting the class of the pairs into ones achieving~$\lim_{t \to \infty} (x(-t ) - x' (-t )) = 0$, we obtain~\eqref{cost-_lb} for any~$(x_0, x'_0) \in \bR^n \times \bR^n$. 
\end{IEEEproof}

\subsection{Proof of Theorem~\ref{thm:rcrsys_Gram}}
\begin{IEEEproof}
First,~\eqref{Ric_con} can be rearranged as
\begin{align*}
&P ( \partial f(x) - B B^\top P) +  ( \partial f(x) - B B^\top P)^\top P\\
&\preceq - \begin{bmatrix}
C \\ - B^\top P
\end{bmatrix}^\top \begin{bmatrix}
C \\ - B^\top P 
\end{bmatrix} - \varepsilon P, \; \forall x \in \bR^n.
\end{align*}
Therefore,~$P$ is a GD observability Gramian of the GD RCR~\eqref{rcrsys}.

Next, direct computations with~\eqref{Ric_con} and~\eqref{Ric_ob} yield
\begin{align*}
&(P + Q^{-1}) ( \partial f(x) - B B^\top P )\\
&\hspace{5mm}+ ( \partial f(x) - B B^\top P )^\top (P + Q^{-1})\\
&= P ( \partial f(x) - B B^\top P ) + ( \partial f(x) - B B^\top P )^\top P\\
&\hspace{5mm}+Q^{-1} ( \partial f(x) Q + Q \partial^\top f(x) \\
&\hspace{20mm} - B B^\top  P Q  - Q  P B B^\top ) Q^{-1}\\
&\preceq - C^\top C - P B B^\top P -\varepsilon P \\
&\hspace{5mm}+ Q^{-1} (  - B B^\top + Q C^\top C Q  \\
&\hspace{20mm} -\varepsilon Q - B B^\top P Q - Q P  B B^\top  ) Q^{-1}\\
&= - (P + Q^{-1}) B B^\top (P + Q^{-1}) - \varepsilon (P + Q^{-1}). 
\end{align*}
Multiplying~$(P + Q^{-1})^{-1}$ from both sides leads to
\begin{align}
&( \partial f(x) - B B^\top P ) (P + Q^{-1})^{-1}\nonumber\\
&\hspace{5mm} + (P + Q^{-1})^{-1} ( \partial f(x) - B B^\top P )^\top \nonumber\\
& \preceq  -  B B^\top - \varepsilon (P + Q^{-1})^{-1}.
\label{GDCG_GDRF}
\end{align}
This implies that~$(P + Q^{-1})^{-1}$ is a GD controllability Gramian of the GD RCR~\eqref{rcrsys}.
\end{IEEEproof}

\subsection{Proof of Theorem~\ref{thm:lcrsys_Gram}}
\begin{IEEEproof}
First,~\eqref{Ric_ob} can be rewritten as
\begin{align*}
&( \partial f(x) - Q C^\top C ) Q + Q ( \partial f(x) - Q C^\top C )^\top \\
&\preceq - \begin{bmatrix} 
B & Q C^\top
\end{bmatrix} 
\begin{bmatrix} 
B & Q C^\top
\end{bmatrix}^\top -\varepsilon P.
\end{align*}
Therefore,~$Q$ is a GD controllability Gramian of the GR LCR~\eqref{lcrsys}.

Next, direct computations with~\eqref{Ric_con} and~\eqref{Ric_ob} lead to
\begin{align*}
&(Q + P^{-1} )^{-1} ( \partial f(x) - Q C^\top C ) \\
&\hspace{4mm}+ ( \partial f(x) -  Q C^\top C )^\top (Q + P^{-1} )^{-1} \\
&\preceq  -  C^\top C - \varepsilon (Q + P^{-1} )^{-1} . 
\end{align*}
Therefore,~$(Q + P^{-1} )^{-1}$ is a GD observability Gramian of the GR LCR~\eqref{lcrsys}.
\end{IEEEproof}

\subsection{Proof of Theorem~\ref{thm:clstab}}
\begin{IEEEproof}
After the change of coordinates~$\xi = x_c -x$ for the controller dynamics, the closed-loop system~\eqref{clsys} becomes
\begin{align*}
\left\{\begin{array}{l}
\dot x = f(x) - B B^\top P x - B B^\top P \xi,\\
\dot \xi = f(\xi + x) - f(x) - Q C^\top C \xi.
\end{array}\right.
\end{align*}
This change of coordinates does not affect stability properties, and thus we analyze this system.

First, we consider~$\xi$. By using the path~$s\xi + x$, we have
\begin{align}
f(\xi + x) - f(x) = \int_0^1 \partial f(s \xi +x) \xi ds.
\label{path_int_z}
\end{align}
Then, it follows from~\eqref{Ric_ob} and~\eqref{path_int_z} that 
\begin{align}
\frac{d }{dt} | \xi |_{Q^{-1}}^2
&= 2 \xi^\top Q^{-1} (f(\xi +x) - f(x) - Q C^\top C \xi )\nonumber\\
&= \int_0^1 \xi^\top ( Q^{-1} \partial f(s \xi +x) + \partial^\top f(s \xi +x)  Q^{-1} ) \xi ds\nonumber\\
&\hspace{4mm} - 2 | C \xi |^2\nonumber\\
&\le \xi^\top Q^{-1}( Q C^\top C Q - B B^\top - \varepsilon Q ) Q^{-1} \xi - 2 | C \xi |^2\nonumber\\
&\le  - \varepsilon | \xi |_{Q^{-1}}^2. \label{con_Q}
\end{align}

Next, we consider~$x$. From Assumption~\ref{asm:eq},~$f(0)= 0$. By using the path~$s x$, we obtain
\begin{align}
f(x) - B B^\top P x 
= \int_0^1 (\partial f(s x) - B B^\top P ) x  ds.
\label{path_int_x*}
\end{align}
Then, it follows from~\eqref{GDCG_GDRF} and~\eqref{path_int_x*} that
\begin{align}
&\frac{d }{dt} |x|_{P + Q^{-1}}^2 \nonumber\\
&=  2x^\top (P + Q^{-1}) (f(x) - B B^\top P x) \nonumber\\
&\hspace{5mm} - 2 x^\top (P + Q^{-1})B B^\top P \xi \nonumber\\
&=  \int_0^1 x^\top ((P + Q^{-1})  (\partial f(s x) - B B^\top P ) \nonumber\\
&\hspace{20mm}+  (\partial f(s x) - B B^\top P )^\top (P + Q^{-1}) ) x  ds \nonumber\\
&\hspace{5mm} - 2 x^\top (P + Q^{-1})B  B^\top P \xi \nonumber\\
&\le -| B^\top (P + Q^{-1}) x|^2 \nonumber\\
&\hspace{5mm} - 2 x^\top (P + Q^{-1})B  B^\top P \xi - \varepsilon |x|_{P + Q^{-1}}^2\nonumber\\
&\le  - \varepsilon |x|_{P + Q^{-1}}^2 + |B^\top P \xi |^2. \label{con_P+Q}
\end{align}

There exists a sufficiently large~$c>0$ satisfying
\begin{align*}
- c \varepsilon Q^{-1} + P B B^\top P \prec 0.
\end{align*}
Therefor, from~\eqref{con_Q} and~\eqref{con_P+Q}, ~$c |\xi^2|_{Q^{-1}} + |x|_{P + Q^{-1}}^2$ is a Lyapunov function for GES at the origin.
\end{IEEEproof}

\section{Proofs for GD $H_\infty$-Balancing}
\subsection{Proof of Theorem~\ref{thm:Hcon}}
\begin{IEEEproof}
In the new coordinates~$\xi = x_c - x$, the closed-loop system becomes
\begin{align*}
\left\{\begin{array}{l}
\dot x = f( x) - B B^\top P_\infty x - B B^\top P_\infty \xi + B w_u,\\
\dot \xi = f(\xi +x) - f(x) -  ( Q_\infty^{-1} -  \gamma^{-2} P_\infty)^{-1} C^\top (C \xi - w_y) \\
\hspace{10mm} + \gamma^{-2} B B^\top P_\infty (x + \xi ) - B w_u\\
z_u = - B^\top P_\infty (x + \xi ), \; z_y = C x.
 \end{array}\right.
\end{align*}
From Assumption~\ref{asm:eq},~$f(0)= 0$, and thus the origin is an equilibrium point when~$w = 0$.

By using the path~$s x$ and~\eqref{Ric_Hcon}, direct computations with~$z_y = Cx$ yield
\begin{align}
& \frac{d}{dt} |x|_{P_\infty}^2 \nonumber\\
&=2 x^\top P_\infty ( f( x) - B B^\top P_\infty x - B B^\top P_\infty \xi + B w_u) \nonumber\\
&\preceq -(1 + \gamma^{-2})  x^\top P_\infty B B^\top P_\infty x \nonumber\\
&\hspace{4mm} - 2 x^\top P_\infty B B^\top P_\infty \xi + 2 x^\top P_\infty B w_u 
 - |C x |^2 - \varepsilon |x|_{P_\infty}^2 \nonumber\\
&= -x^\top P_\infty B B^\top P_\infty x  - 2 x^\top P_\infty B B^\top P_\infty \xi \nonumber\\
&\hspace{4mm}- \gamma^{-2} | B^\top P_\infty x  -  \gamma^2 w_u |^2 + \gamma^2 | w_u |^2
 - |z_y |^2 - \varepsilon |x|_{P_\infty}^2. \label{ineq:Pinf}
\end{align}

Next, we consider  $|\xi|_{\gamma^2 Q_\infty^{-1} - P_\infty}^2$ that is positive definite because of item III). By using the path~$s\xi +x$, direct computations with~\eqref{def_R} and~\eqref{Ric_Hob} yield
\begin{align*}
&\frac{d}{dt} |\xi|_{\gamma^2 Q_\infty^{-1} - P_\infty}^2\\
&=2 \xi^\top (\gamma^2 Q_\infty^{-1} - P_\infty) (f(\xi +x) - f(x))\\
&\hspace{4mm}-2 \gamma^2 \xi^\top  C^\top (C \xi - w_y) \\
&\hspace{4mm} +2\gamma^{-2}  \xi^\top (\gamma^2 Q_\infty^{-1} - P_\infty) B (B^\top P_\infty (x +  \xi) - \gamma^2 w_u)\\
&\preceq  \gamma^2 \xi^\top( (1-\gamma^{-2}) C^\top C -  Q_\infty^{-1} B B^\top Q_\infty^{-1}- \varepsilon  Q_\infty^{-1}) \xi\\
&\hspace{4mm}+ \xi^\top ( R_\infty (x) + C^\top C + \varepsilon P_\infty - (1-\gamma^{-2}) P_\infty B B^\top P_\infty) \xi\\
&\hspace{4mm}- \xi^\top R_\infty (x) \xi  -2 \gamma^2 \xi^\top  C^\top (C \xi - w_y) \\
&\hspace{4mm} +2\gamma^{-2}  \xi^\top (\gamma^2 Q_\infty^{-1} - P_\infty) B (B^\top P_\infty (x +  \xi) - \gamma^2 w_u)\\
&=  - \xi^\top  P_\infty B B^\top P_\infty \xi - \gamma^2 \xi^\top Q_\infty^{-1} B B^\top Q_\infty^{-1} \xi \\
&\hspace{4mm} +2 \xi^\top Q_\infty^{-1} B B^\top P_\infty \xi - \gamma^{-2}  \xi^\top P_\infty B B^\top P_\infty \xi\\
&\hspace{4mm} +2\gamma^{-2}  \xi^\top (\gamma^2 Q_\infty^{-1} - P_\infty) B (B^\top P_\infty x - \gamma^2 w_u)\\
&\hspace{4mm}- \gamma^2 \xi^\top C^\top C \xi + 2 \gamma^2 \xi^\top  C^\top w_y 
- \varepsilon  |\xi|_{\gamma^2 Q_\infty^{-1} - P_\infty}^2 \\
&=  - \xi^\top P_\infty B B^\top P_\infty \xi  - \gamma^{-2} | B^\top ( \gamma^2  Q_\infty^{-1} - P_\infty ) \xi |^2 \\
&\hspace{4mm} +2\gamma^{-2}  \xi^\top (\gamma^2 Q_\infty^{-1} - P_\infty) B (B^\top P_\infty x - \gamma^2 w_u)\\
&\hspace{4mm}- \gamma^2 \xi^\top C^\top C \xi + 2 \gamma^2 \xi^\top  C^\top w_y - \varepsilon  |\xi|_{\gamma^2 Q_\infty^{-1} - P_\infty}^2 \\
&\preceq - \xi^\top P_\infty B B^\top P_\infty \xi   + \gamma^{-2} | B^\top P_\infty x - \gamma^2 w_u|^2\\
&\hspace{4mm}+\gamma^2 |w_y|^2 - \varepsilon  |\xi|_{\gamma^2 Q_\infty^{-1} - P_\infty}^2.
\end{align*}

By summing up the above two inequalities with~$z_u = - B^\top P_\infty (x+ \xi)$, we obtain 
\begin{align*}
& \frac{d}{dt} \left( |x|_{P_\infty}^2 + |\xi|_{\gamma^2 Q_\infty^{-1} - P_\infty}^2 \right)
 +  \varepsilon  \left(|x|_{P_\infty}^2 + |\xi|_{\gamma^2 Q_\infty^{-1} - P_\infty}^2 \right)\\
&\preceq  -x^\top P_\infty B B^\top P_\infty x  - 2 x^\top P_\infty B B^\top P_\infty \xi \\
&\hspace{4mm}- \xi^\top P_\infty B B^\top P_\infty \xi + \gamma^2 ( | w_u |^2 +  |w_y|^2) - |z_y |^2 \\
&= - ( |z_u |^2 + |z_y |^2 ) + \gamma^2 ( | w_u |^2 +  |w_y|^2).
\end{align*}
By taking the time integration over the interval~$[0, t]$, one can conclude that item 1) of Problem~\ref{prob:Hcon} is solved. Moreover, item~2) is solved when~$\varepsilon > 0$ for~$w = 0$.
\end{IEEEproof}

\subsection{Proof of Theorem~\ref{thm:small_gamma}}
\begin{IEEEproof}
First, $\varepsilon = \varepsilon_1+2\varepsilon_2$, \eqref{Ric_Hcon} and~\eqref{alpha_P} yield 
\begin{align*}
&P_\infty \partial f(x) + \partial^\top f(x) P_\infty - (\beta^2 - \alpha) P_\infty B B^\top P_\infty\\
& + C^\top C + (\varepsilon_1+\varepsilon_2) P_\infty \preceq 0,
\end{align*}
and thus~$P_\infty$ satisfies~\eqref{Ric_Hcon} for~$\hat \beta := \sqrt{\beta^2 - \alpha}>0$.
Next, from~\eqref{def_R}, one can confirm that its left-hand side is lower bounded on~$-\bar R_\infty (x)$. In addition, from~\eqref{Ric_Hcon}, $Q_\infty$ satisfies
\begin{align*}
&\partial f(x) Q_\infty + Q_\infty \partial^\top f(x) - \beta^2 Q_\infty C^\top C Q_\infty 
+ B B^\top\\
& + \varepsilon Q_\infty - \varepsilon_2 \gamma^{-2} Q_\infty P_\infty Q_\infty
 \preceq - \gamma^{-2} Q_\infty \bar R_\infty (x) Q_\infty.
\end{align*}
Item 3) of Theorem~\ref{thm:Hcon} and~\eqref{alpha_Q} imply that~$Q_\infty$ satisfies~\eqref{Ric_Hob} for~$\hat \beta$. Finally, it follows that~$\hat \beta = \sqrt{1 - \bar \gamma^{-2}}$.
\end{IEEEproof}

\subsection{Proof of Lemma~\ref{lem:Hcon_gain}}
\begin{IEEEproof}
Let~$\xi_r:=x_{cr}-x_r$ and~$V_r(x_r,\xi_r) := |x_r|_{\Pi_1}^2 + |\xi_r|_{\gamma^2 \Pi_1^{-1} - \Pi_1}^2$. Then, in a similar manner as the proof of Theorem~\ref{thm:Hcon}, it follows that
\begin{align*}
&\frac{d}{dt} V_r(x_r,\xi_r) + \varepsilon V_r(x_r,\xi_r)\\
&\le  - (|z_{ur}|^2 +|z_{yr}|^2) + \gamma^2 (|w_{ur}|^2 +|w_{yr}|^2)\\
&= - (|u_r - w_{ur} |^2 +|\beta^{-1} \bar y_r|^2) + \gamma^2 (|w_{ur}|^2 +| \beta^{-1} \bar w_{yr}|^2),
\end{align*}
where in the last equality, recall the definitions of signals in~\eqref{rgsys} and~\eqref{signals}.
In a similar manner as the proof of~\cite[Lemma 9 (ii)]{PF:97}, it is possible to show for~$\bar V_r(x_r,\xi_r):=(1+\gamma^{-1} \beta)V_r(x_r,\xi_r)$ that
\begin{align}
&\frac{d}{dt} \bar V_r(x_r,\xi_r) + \varepsilon \bar V_r(x_r,\xi_r) \nonumber\\
&\le - (|u_r|^2 +|\bar y_r|^2) + (1+\gamma \beta^{-1})^2 | \bar w_r |^2,
\label{gain_pre1}
\end{align}
where again recall the definitions of~$\bar w_r$ in~\eqref{signals}.

On the other hands, let~$\bar \Pi_1:= \beta^2 \Pi_1$. For the system~$M_r^{-1}$ in~\eqref{Mrsys}, the direct computation with~\eqref{Ric_Hcon} yields
\begin{align}
& \frac{d}{dt} |x_r|_{\bar \Pi_1}^2 + \varepsilon |x_r|_{\bar \Pi_1}^2 \nonumber\\
&\le x_r^\top \bar \Pi_1 B_1 B_1^\top \bar \Pi_1 x_r + 2 x_r^\top \bar \Pi_1 B_1 u_r - \beta^2 |C_1 x_r |^2\nonumber\\
&= |B_1^\top \bar \Pi_1 x_r + u_r|^2 - |u_r|^2 - \beta^2 |C_1 x_r |^2\nonumber\\
&= |v|^2 - (|u_r|^2 + |\bar y_r |^2),
\label{gain_pre2}
\end{align}
where in the last equality, the definitions of $v$ in~\eqref{Mrsys}, $y_r$ in~\eqref{rgsys}, and~$\bar y_r$ in~\eqref{signals} are used. 

Now, define
\begin{align*}
&W_r(x_r,\xi_r) = \bar V(x_r,\xi_r) - |x_r|_{\bar \Pi_1}^2\\
&\hspace{5mm}=  (1+\gamma^{-1} \beta - \beta^2 )  |x_r|_{\Pi_1}^2
+ (1+\gamma^{-1} \beta)  |\xi_r|_{\gamma^2 \Pi_1^{-1} - \Pi_1}^2.
\end{align*}
Note that~$1+\gamma^{-1} \beta - \beta^2 > 0$. Therefore,~$W_r(x_r,\xi_r)$ is positive definite. We compute its time derivative. Since~\eqref{ineq:Pinf} and~\eqref{gain_pre2} are based on the same inequality~\eqref{Ric_Hcon}, by decomposing the time derivative of~$(1+\gamma^{-1} \beta - \beta^2 )  |x_r|_{\Pi_1}^2$ into two parts, we obtain from~\eqref{gain_pre1} and~\eqref{gain_pre2},
\begin{align}
& \frac{d}{dt} W_r(x_r,\xi_r) + \varepsilon W_r(x_r,\xi_r) \nonumber\\
&\le -   |v|^2  + (1+\gamma \beta^{-1})^2 | \bar w_r |^2 .
\label{gain_w_v}
\end{align}
Recall~$\bar w_r = \bar e + \bar w$ in~\eqref{signals}. By taking its time integration over~$[0,\infty)$ and using the triangular inequality, we obtain the statement of this lemma.
\end{IEEEproof}

\subsection{Proof of Theorem~\ref{thm:clstab_rconH}}
\begin{IEEEproof}
[Proof of item 1)]
By using Lemma~\ref{lem:Hcon_gain}, Corollary~\ref{error_rcrHsys:cor}, and the counterpart of Theorem~\ref{thm:Hcon} to the reduced-order closed-loop system~$\Gamma (G_r,K_r)$, item~1) can be shown by following a similar procedure as the proof of~\cite[Theorem 4]{PF:97}, where~$\varepsilon$ in~\cite{PF:97} is~$\rho_r/\beta$ of this paper. 
\end{IEEEproof}

\begin{IEEEproof}
[Proof of item 2)]
We consider the perturbation~$\Delta$. The modification of~\eqref{pre:clstab_rconH} to the GD RCR~\eqref{rcrHsys}  is
\begin{align}
&\underline{c}(|x(t)|^2 + |x_r(t)|^2) + \varepsilon \int_0^t \underline{c}(|x(\tau)|^2 + |x_r(\tau )|^2) d\tau\nonumber\\
&+ \int_0^t | \bar e (\tau )|^2 d\tau \nonumber\\
&\le \left( 1 + \frac{1}{\theta} \right) \overline{c}(|x(0)|^2 + |x_r(0)|^2) +  \rho_r^2 (1 + \theta) \int_0^\tau | v (t)|^2 dt
\label{perturb_bd}
\end{align}
for any~$\theta > 0$, where the initial states of~$r+1$ to~$n-1$-dimensional reduced-order models are zero.

Next, we study the feedback interconnection of~$\Gamma (G_r,K_r): \bar e \to v$ and~$\Delta:v \to \bar e$ for~$\bar w=0$. Let~$V(x_r, \xi_r, x, x_r):= \rho_r^2 (1 + \theta ) W_r(x_r,\xi_r) + \underline{c}(|x|^2 + |x_r|^2)$ for $W_r(x_r,\xi_r)$ in~\eqref{gain_w_v}. Then, the two inequalities~\eqref{gain_w_v} with~$\bar w_r = \bar e$ and~\eqref{perturb_bd} yield
\begin{align*}
& V(x_r(\tau ), \xi_r(\tau ), x(\tau ), x_r(\tau ))\\
&\hspace{4mm} + \varepsilon \int_0^\tau V(x_r(t), \xi_r(t), x(t), x_r(t )) dt\\
&\hspace{4mm}+ \left( 1- \rho_r^2 (1 + \theta ) (1+\gamma \beta^{-1})^2 \right) \int_0^\tau | \bar e (t) |^2 dt\\ 
&\le \rho_r^2 (1 + \theta ) W_r(x_r(0),\xi_r(0))\\
&\hspace{4mm} + (1 + 1/\theta ) \overline{c}(|x(0)|^2 + |x_r(0)|^2).  
\end{align*}
From item II), $1- \rho_r^2 (1+\gamma \beta^{-1})^2>0$, which implies that there exists a sufficiently small~$\theta > 0$ such that 
\begin{align*}
1- \rho_r^2 (1 + \theta ) (1+\gamma \beta^{-1})^2 > 0.
\end{align*}
Also, there exists~$\hat c>0$ such that
\begin{align*}
&\rho_r^2 (1 + \theta ) W_r(x_r,\xi_r) + (1 + 1/\theta ) \overline{c}(|x|^2 + |x_r|^2)\\
&\le \hat c V(x_r, \xi_r, x, x_r).
\end{align*}
Therefore, we obtain
\begin{align*}
& V(x_r(\tau ), \xi_r(\tau ), x(\tau ), x_r(\tau ))\\
&\hspace{4mm} + \varepsilon \int_0^\tau V(x_r(t), \xi_r(t), x(t), x_r(t )) dt\\
&\le \hat c V(x_r(0), \xi_r(0), x(0), x_r(0)), \; \forall t \in \bR_+.
\end{align*}
From the Gronwall inequality~\cite[Lemma A.1]{Khalil:02},  the interconnected system of~$\Gamma (G_r,K_r)$ and~$\Delta$ is GES at the origin. When~$x_r(0) = 0$, this implies the GES at the origin of the closed-loop system~$\Gamma (G,K_r)$ with the states~$x$ and~$x_{cr}$.
\end{IEEEproof}

\bibliographystyle{IEEEtran} 
\bibliography{RefGDB}

% Generated by IEEEtran.bst, version: 1.13 (2008/09/30)
\begin{thebibliography}{10}
\providecommand{\url}[1]{#1}
\csname url@samestyle\endcsname
\providecommand{\newblock}{\relax}
\providecommand{\bibinfo}[2]{#2}
\providecommand{\BIBentrySTDinterwordspacing}{\spaceskip=0pt\relax}
\providecommand{\BIBentryALTinterwordstretchfactor}{4}
\providecommand{\BIBentryALTinterwordspacing}{\spaceskip=\fontdimen2\font plus
\BIBentryALTinterwordstretchfactor\fontdimen3\font minus
  \fontdimen4\font\relax}
\providecommand{\BIBforeignlanguage}[2]{{%
\expandafter\ifx\csname l@#1\endcsname\relax
\typeout{** WARNING: IEEEtran.bst: No hyphenation pattern has been}%
\typeout{** loaded for the language `#1'. Using the pattern for}%
\typeout{** the default language instead.}%
\else
\language=\csname l@#1\endcsname
\fi
#2}}
\providecommand{\BIBdecl}{\relax}
\BIBdecl

\bibitem{LMG:02}
S.~Lall, J.~E. Marsden, and S.~Glava{\v{s}}ki, ``A subspace approach to
  balanced truncation for model reduction of nonlinear control systems,''
  \emph{Int. J. Robust Nonlinear Cont.}, vol.~12, no.~6, pp. 519--535, 2002.

\bibitem{FT:08}
K.~Fujimoto and D.~Tsubakino, ``Computation of nonlinear balanced realization
  and model reduction based on {T}aylor series expansion,'' \emph{Sys. Cont.
  Lett.}, vol.~57, no.~4, pp. 283--289, 2008.

\bibitem{HLB:12}
P.~Holmes, J.~L. Lumley, G.~Berkooz, and C.~W. Rowley, \emph{Turbulence,
  Coherent Structures, Dynamical Systems and Symmetry}.\hskip 1em plus 0.5em
  minus 0.4em\relax Cambridge: Cambridge University Press, 2012.

\bibitem{Kashima:16}
K.~Kashima, ``Noise response data reveal novel controllability {G}ramian for
  nonlinear network dynamics,'' \emph{Sci. Rep.}, vol.~6, no. 27300, 2016.

\bibitem{KS:21}
Y.~Kawano and J.~M.~A. Scherpen, ``Empirical differential {G}ramians for
  nonlinear model reduction,'' \emph{Automatica}, vol. 127, p. 109534, 2021.

\bibitem{Scherpen:93}
J.~M.~A. Scherpen, ``Balancing for nonlinear systems,'' \emph{Sys. Cont.
  Lett.}, vol.~21, no.~2, pp. 143--153, 1993.

\bibitem{FS:05}
K.~Fujimoto and J.~M.~A. Scherpen, ``Nonlinear input-normal realizations based
  on the differential eigenstructure of {H}ankel operators,'' \emph{IEEE Trans.
  Autom. Control}, vol.~50, no.~1, pp. 2--18, 2005.

\bibitem{BVS:14}
B.~Besselink, N.~van~de Wouw, J.~M.~A. Scherpen, and H.~Nijmeijer, ``Model
  reduction for nonlinear systems by incremental balanced truncation,''
  \emph{IEEE Trans. Autom. Control}, vol.~59, no.~10, pp. 2739 -- 2753, 2014.

\bibitem{KS:17}
Y.~Kawano and J.~M.~A. Scherpen, ``Model reduction by differential balancing
  based on nonlinear {H}ankel operators,'' \emph{IEEE Trans. Autom. Control},
  vol.~62, no.~7, pp. 3293--3308, 2017.

\bibitem{KS:19}
------, ``Structure preserving truncation of nonlinear port {H}amiltonian
  systems,'' \emph{IEEE Trans. Autom. Control}, vol.~62, no.~2, 2019.

\bibitem{Astolfi:10}
A.~Astolfi, ``Model reduction by moment matching for linear and nonlinear
  systems,'' \emph{IEEE Trans. Autom. Control}, vol.~55, no.~10, pp.
  2321--2336, 2010.

\bibitem{BVS:13}
B.~Besselink, N.~van~de Wouw, J.~M.~A. Scherpen, and H.~Nijmeijer,
  ``Generalized incremental balanced truncation for nonlinear systems,''
  \emph{Proc. 52nd IEEE Conf. Dec. Cont.}, pp. 5552--5557, 2013.

\bibitem{KS:15}
Y.~Kawano and J.~M.~A. Scherpen, ``Model reduction by generalized differential
  balancing,'' in \emph{Mathematical Control Theory I}, M.~K. Camlibel, A.~A.
  Julius, R.~Pasumarthy, and J.~M.~A. Scherpen, Eds.\hskip 1em plus 0.5em minus
  0.4em\relax Springer-Verlag, pp. 349-362, 2015.

\bibitem{SA:17}
G.~Scarciotti and A.~Astolfi, ``Data-driven model reduction by moment matching
  for linear and nonlinear systems,'' \emph{Automatica}, vol.~79, pp. 340--351,
  2017.

\bibitem{KBS:20}
Y.~Kawano, B.~Besselink, J.~M.~A. Scherpen, and M.~Cao, ``Data-driven model
  reduction of monotone systems by nonlinear {DC} gains,'' \emph{IEEE Trans.
  Autom. Control}, vol.~65, no.~5, pp. 2094--2167, 2020.

\bibitem{JS:83}
E.~Jonckheere and L.~Silverman, ``A new set of invariants for linear systems --
  application to reduced order compensator design,'' \emph{IEEE Trans. Autom.
  Control}, vol.~28, no.~10, pp. 953--964, 1983.

\bibitem{OM:89}
R.~Ober and D.~Mc{F}arlane, ``Balanced canonical forms for minimal systems: A
  normalized coprime factor approach,'' \emph{Lin. Alg. and its Applications},
  vol. 122, pp. 23--64, 1989.

\bibitem{MG:91}
D.~Mustafa and K.~Glover, ``Controller reduction by
  $\mathcal{H}_\infty$-balanced truncation,'' \emph{IEEE Trans. Autom.
  Control}, vol.~36, no.~6, pp. 668--682, 1991.

\bibitem{OA:12}
G.~Obinata and B.~D.~O. Anderson, \emph{Model reduction for control system
  design}.\hskip 1em plus 0.5em minus 0.4em\relax Springer Science \& Business
  Media, 2012.

\bibitem{PF:97}
L.~Pavel and F.~Fairman, ``Controller reduction for nonlinear plants -- an
  $l_2$ approach,'' \emph{Int. J. Robust Nonlinear Cont.}, vol.~7, no.~5, pp.
  475--505, 1997.

\bibitem{YW:01}
C.-F. Yung and H.-S. Wang, ``${H}^\infty$ controller reduction for nonlinear
  systems,'' \emph{Automatica}, vol.~37, no.~11, pp. 1797--1802, 2001.

\bibitem{LS:98}
W.~Lohmiller and J.-J.~E. Slotine, ``On contraction analysis for non-linear
  systems,'' \emph{Automatica}, vol.~34, no.~6, pp. 683--696, 1998.

\bibitem{FS:14}
F.~Forni and R.~Sepulchre, ``A differential {L}yapunov framework for
  contraction anlaysis,'' \emph{IEEE Trans. Autom. Control}, vol.~59, no.~3,
  pp. 614--628, 2014.

\bibitem{KBC:20}
Y.~Kawano, B.~Besselink, and M.~Cao, ``Contraction analysis of monotone systems
  via separable functions,'' \emph{IEEE Trans. Autom. Control}, vol.~65, no.~8,
  pp. 3486--3501, 2020.

\bibitem{SB:04}
J.~W. Simpson-Porco and F.~Bullo, ``Contraction theory on riemannian
  manifolds,'' \emph{Sys. Cont. Lett.}, vol.~65, pp. 74--80, 2014.

\bibitem{AJP:16}
V.~Andrieu, B.~Jayawardhana, and L.~Praly, ``Transverse exponential stability
  and applications,'' \emph{IEEE Trans. Autom. Control}, vol.~61, no.~11, pp.
  3396--3411, 2016.

\bibitem{Sontag:10}
E.~D. Sontag, ``Contractive systems with inputs,'' in \emph{Perspectives in
  Mathematical System Theory, Control, and Signal Processing}.\hskip 1em plus
  0.5em minus 0.4em\relax Springer, 2010, pp. 217--228.

\bibitem{KCS:21}
Y.~Kawano, C.~K. Kosaraju, and J.~M.~A. Scherpen, ``Krasovskii and shifted
  passivity based control,'' \emph{IEEE Trans. Autom. Control}, vol.~66,
  no.~10, pp. 4926--4932, 2021.

\bibitem{KH:21}
Y.~Kawano and Y.~Hosoe, ``Contraction analysis of discrete-time stochastic
  systems,'' \emph{arXiv preprint arXiv:2106.05635}, 2021.

\bibitem{Angeli:02}
D.~Angeli, ``A {L}yapunov approach to incremental stability properties,''
  \emph{IEEE Trans. Autom. Control}, vol.~47, no.~3, pp. 410--421, 2002.

\bibitem{GA:04}
S.~Gugercin and A.~C. Antoulas, ``A survey of model reduction by balanced
  truncation and some new results,'' \emph{Int. J. of Control}, vol.~77, no.~8,
  pp. 748--766, 2004.

\bibitem{BM:08}
F.~Blanchini and S.~Miani, \emph{Set-Theoretic Methods in Control}.\hskip 1em
  plus 0.5em minus 0.4em\relax Birkh\"{a}user, 2008.

\bibitem{Antoulas:05}
A.~C. Antoulas, \emph{Approximation of Large-Scale Dynamical Systems}.\hskip
  1em plus 0.5em minus 0.4em\relax Philadelphia: SIAM, 2005.

\bibitem{Moore:81}
B.~Moore, ``Principal component analysis in linear systems: Controllability,
  observability, and model reduction,'' \emph{IEEE Trans. Autom. Control},
  vol.~26, no.~1, pp. 17--32, 1981.

\bibitem{Khalil:02}
H.~Khalil, \emph{Nonlinear Systems}, 3rd~ed.\hskip 1em plus 0.5em minus
  0.4em\relax Prentice Hall, 2002.

\bibitem{HSK:92}
J.~Hauser, S.~Sastry, and P.~Kokotovic, ``Nonlinear control via approximate
  input-output linearization: The ball and beam example,'' \emph{IEEE Trans.
  Autom. Control}, vol.~37, no.~3, pp. 392--398, 1992.

\bibitem{SS:94}
J.~M.~A. Scherpen and A.~J. van~der Schaft, ``Normalized coprime factorizations
  and balancing for unstable nonlinear systems,'' \emph{Int. J. Control},
  vol.~60, no.~6, pp. 1193--1222, 1994.

\bibitem{DGK:89}
J.~C. Doyle, K.~Glover, P.~P. Khargonekar, and B.~A. Francis, ``State-space
  solutions to standard $\mathcal{H}_2$ and $\mathcal{H}_\infty$ control
  problems,'' \emph{IEEE Trans. Autom. Control}, vol.~34, no.~8, pp. 831--847,
  1989.

\bibitem{LD:94}
W.-M. Lu and J.~C. Doyle, ``$\mathcal{H}_\infty$ control of nonlinear systems
  via output feedback: controller parameterization,'' \emph{IEEE Trans. Autom.
  Control}, vol.~39, no.~12, pp. 2517--2521, 1994.

\bibitem{Banach:22}
S.~Banach, ``Sur les op\'erations dans les ensembles abstraits et leur
  application aux \'equations int\'egrales,'' \emph{Fund. Math.}, vol.~3,
  no.~1, pp. 133--181, 1922.

\bibitem{Palais:59}
R.~S. Palais, ``Natural operations on differential forms,'' \emph{Trans. Amer.
  Math. Soc.}, vol.~92, no.~1, pp. 125--141, 1959.

\end{thebibliography}

\end{document}